\documentclass[ee,12pt,a4paper,oneside]{genthesis}
\usepackage[linktoc=page]{hyperref}
\usepackage{color}
\hypersetup{
	colorlinks=true,
	linkcolor=blue, 
    citecolor=blue,
    filecolor=blue,
    linkcolor=blue,
    urlcolor=blue
}
\usepackage{graphicx}

\usepackage{caption}

\usepackage{multirow}
\usepackage{colortbl}

\usepackage{fancyhdr}
\usepackage{mleftright}

\usepackage{multicol}
\usepackage{booktabs}
\usepackage{csquotes}
\usepackage{exscale,relsize}

\usepackage{footnotebackref}
\usepackage[utf8]{inputenc}

\usepackage[greek,english]{babel}
	\addto{\captionsenglish}{\renewcommand{\bibname}{References}}	
\usepackage{setspace}
\usepackage{times}
\usepackage{epstopdf}
\usepackage{algorithm2e}
\usepackage{algorithmicx}

\usepackage{tikz}
\usepackage{siunitx}
\usepackage{eurosym}
\usepackage{multirow}

\usepackage{amsmath}
\usepackage{amssymb}

\usepackage{amsthm}
\usepackage{listings}
\usepackage [short]{optidef}
\usetikzlibrary{shapes,arrows}
\makeatletter
\newsavebox\myboxA
\newsavebox\myboxB
\newlength\mylenA

\newcommand*\xoverline[2][0.75]{%
    \sbox{\myboxA}{$\m@th#2$}%
    \setbox\myboxB\null
    \ht\myboxB=\ht\myboxA%
    \dp\myboxB=\dp\myboxA%
    \wd\myboxB=#1\wd\myboxA
    \sbox\myboxB{$\m@th\overline{\copy\myboxB}$}
    \setlength\mylenA{\the\wd\myboxA}
    \addtolength\mylenA{-\the\wd\myboxB}%
    \ifdim\wd\myboxB<\wd\myboxA%
       \rlap{\hskip 0.5\mylenA\usebox\myboxB}{\usebox\myboxA}%
    \else
        \hskip -0.5\mylenA\rlap{\usebox\myboxA}{\hskip 0.5\mylenA\usebox\myboxB}%
    \fi}
\makeatother

\DeclareRobustCommand{\rchi}{{\mathpalette\irchi\relax}}
\newcommand{\irchi}[2]{\raisebox{\depth}{$#1\chi$}} 

\newcommand{\vecx}{\mbox{\boldmath $x$}}
\newcommand{\vecM}{\mbox{\boldmath $M$}}
\newcommand{\vecD}{\mbox{\boldmath $\Delta$}}
\newcommand{\vecn}{\mbox{\boldmath $n$}}
\newcommand{\vecrr}{\mbox{\boldmath $\varrho$}}
\newcommand{\veclt}{\mbox{\boldmath $l$}}
\newcommand{\vecA}{\mbox{\boldmath $A$}}
\newcommand{\vecAbar}{\mbox{\boldmath $\bar{A}$}}
\newcommand{\vecG}{\mbox{\boldmath $\Gamma$}}
\newcommand{\veck}{\mbox{\boldmath $\varkappa$}}
\newcommand{\vecf}{\mbox{\boldmath $f$}}
\newcommand{\vece}{\mbox{\boldmath $e$}}

\newcommand{\vecv}{\mbox{\boldmath $v$}}
\newcommand{\vecu}{\mbox{\boldmath $u$}}
\newcommand{\vecc}{\mbox{\boldmath $c$}}

\newcommand{\vecC}{\mbox{\boldmath $C$}}
\newcommand{\vecy}{\mbox{\boldmath $y$}}
\newcommand{\vecl}{\mbox{\boldmath $\lambda$}}
\newcommand{\vecY}{\mbox{\boldmath $Y$}}
\newcommand{\vecw}{\mbox{\boldmath $w$}}
\newcommand{\vecp}{\mbox{\boldmath $\pi$}}
\newcommand{\vecz}{\mbox{\boldmath $z$}}
\newcommand{\vecq}{\mbox{\boldmath $q$}}
\newcommand{\vecones}{\mbox{\boldmath $1$}}

\newcommand{\vecd}{\mbox{\boldmath $\delta$}}

\newcommand{\vecwhat}{\mbox{\boldmath $\hat{w}$}}
\newcommand{\vecm}{\mbox{\boldmath $\mu$}}
\newcommand{\vecmhat}{\mbox{\boldmath $\hat{\mu}$}}
\newcommand{\vecytilde}{\mbox{\boldmath $\tilde{y}$}}
\newcommand{\vecFtilde}{\mbox{\boldmath $\tilde{F}$}}
\newcommand{\vecS}{\mbox{\boldmath $\varSigma$}}
\newcommand{\vecShat}{\mbox{\boldmath $\hat{\varSigma}$}}

\newcommand{\emptypage}{\null
\thispagestyle{empty}%
\addtocounter{page}{-1}%
\newpage}
\newcommand{\makegreektitlepage}{{
\greektext
  \thispagestyle{empty}
  \vspace*{0in}
  \begin{center}%
  	{ΠΟΛΥΤΕΧΝΕΙΟ ΚΡΗΤΗΣ}\\
	{ΣΧΟΛΗ ΜΗΧΑΝΙΚΩΝ ΠΑΡΑΓΩΓΗΣ ΚΑΙ ΔΙΟΙΚΗΣΗΣ}\\
	\par%
  	\vspace{1em}%
  	\includegraphics[scale=0.7]{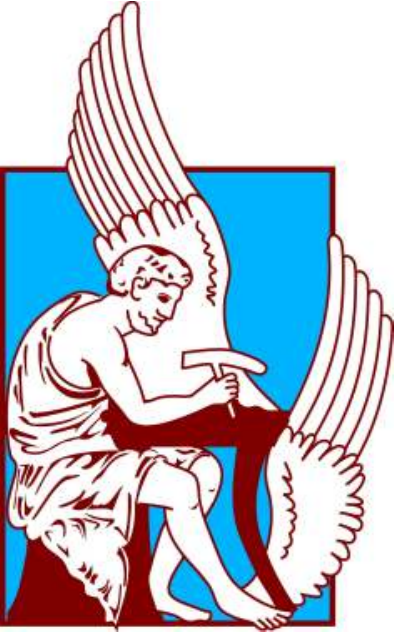}\par%
  	\vspace{1em}%
  \end{center}
  \begin{center}
   \large {Μεθοδολογίες Ευσταθούς Βελτιστοποίησης Επενδυτικών Χαρτοφυλακίων: Μια Υπολογιστική Συγκριτική Ανάλυση}\\
  \end{center}
  \vspace{.2in}
  \begin{center}
	{Αντώνιος Γεωργαντάς}
  \end{center}
  \vspace{.2in}
  \begin{center}
    {ΔΙΑΤΡΙΒΗ ΓΙΑ ΤΗ ΜΕΡΙΚΗ ΕΚΠΛΗΡΩΣΗ}\\
    {ΤΩΝ ΑΠΑΙΤΗΣΕΩΝ ΓΙΑ ΤΟ ΜΔΕ}\\
    {ΣΤΗΝ ΕΠΙΧΕΙΡΗΣΙΑΚΗ ΕΡΕΥΝΑ}
  \end{center}
  \vspace{.2in}
  \begin{center}
    {ΙΟΥΝΙΟΣ 2018}
  \end{center}  
  \vspace{.5in}
  \begin{center}
     {ΕΞΕΤΑΣΤΙΚΗ ΕΠΙΤΡΟΠΗ}\\
	 {Καθηγητής Μιχάλης Δούμπος (ΜΠΔ), Επιβλέπων}\\
	 {Καθηγητής  Κωνσταντίνος Ζοπουνίδης (ΜΠΔ)}\\
	 {Καθηγητής Φώτιος Πασιούρας (ΜΠΔ)}
  \end{center}
  \clearpage
  }}
\renewcommand\thepart         {\Roman{part}}
\renewcommand\thechapter      {\arabic{chapter}}
\renewcommand\thesection      {\thechapter.\arabic{section}}
\renewcommand\thesubsection   {\thesection.\arabic{subsection}}
\renewcommand\thesubsubsection{\thesubsection.\arabic{subsubsection}}
\renewcommand\theparagraph    {\thesubsubsection.\arabic{paragraph}}

\DeclareTextFontCommand{\emph}{\em}
\DeclareSIUnit{\pers}{pers}
\DeclareSIUnit{\EUR}{\text{\euro}}
\DeclareRobustCommand{\officialeuro}{%
  \ifmmode\expandafter\text\fi
  {\fontencoding{U}\fontfamily{eurosym}\selectfont e}}

\title{Robust Optimization Approaches for Portfolio Selection: A Computational and Comparative Analysis}
\author{Antonios Georgantas}
\submitdate{June 2018}
\advisor{Professor Michalis Doumpos}

\begin{document}

\def\mathbi#1{\textbf{\em #1}}

\maketitle
\emptypage
\makegreektitlepage

\chapter*{Abstract}

The field of portfolio selection is an active research topic, which combines elements and methodologies from various fields, such as optimization, decision analysis, risk management, data science, forecasting, etc. The modeling and treatment of deep uncertainties for the future asset returns is a major issue for the success of analytical portfolio selection models. Recently, robust optimization (RO) models have attracted a lot of interest in this area. 
RO provides a computationally tractable framework for portfolio optimization based on relatively general assumptions on the probability distributions of the uncertain risk parameters. Thus, RO extends the framework of traditional linear and non-linear models (e.g., the well-known mean-variance model), incorporating uncertainty through a formal and analytical approach into the modeling process. Robust counterparts of existing models can be considered as worst-case re-formulations as far as deviations of the uncertain parameters from their nominal values are concerned.
Although several RO models have been proposed in the literature focusing on various risk measures and different types of uncertainty sets about asset returns, analytical empirical assessments of their performance have not been performed in a comprehensive manner. The objective of this study is to fill in this gap in the literature. More specifically, we consider different types of RO models based on popular risk measures and conduct an extensive comparative analysis of their performance using data from the US market during the period 2005-2016. For the analysis, three different robust versions of the mean-variance model are considered, together with two other robust models for conditional value-at-risk and the omega ratio. The robust versions are compared against standard (non-robust) models through various portfolio performance metrics, focusing on out-of-sample results. The analysis is based on a rolling window approach.

\chapter*{Acknowledgments}

By completing the present dissertation, I would like to thank Professor Doumpos for giving me the opportunity to work under his guidance in the field of Financial Engineering, as also for the patience and perseverance he has shown throughout this period. Professor Doumpos provided me with everything he deemed  important in order to grasp effectively the notions affiliated with the Thesis, with respect to what we were opting to accomplish. The interaction was more than satisfying and this acted as the foundation for many of the methodologies I explain in the main body of the Thesis. Furthermore, I am grateful to my family and the friends close to me for their support during this period.

\tableofcontents

\chapter{Introduction}
\label{ch:intro}

An area of science which constantly admits novel research is financial engineering. Financial engineering is a multi-disciplinary field involving financial theory, the tools of mathematics and the practice of programming. One popular field of research in this area concerns the portfolio selection and management. The seminal approach of Markowitz during 1950s focusing on the development of a mean-variance model through quadratic mathematical modeling tools, was the cornerstone in this area and altered the philosophy in the financial domain. Although the mean-variance model is extensively employed by practitioners, it has several impractical aspects. One discrete element is that Markowitz’s framework considers the first two moments of the distribution and therefore implies that the underlying asset returns are normally distributed~\cite{fabozzi2007robust}. In addition to unrealistic portfolio weights, one of the major discrepancies with the mean-variance model is the high sensitivity of the parameter estimations to small changes in the inputs. This doesn’t seem to be that appealing to the broader community and a different scheme which accounts for less restrictive and more realistic assumptions needs to be formulated. An idea which could alleviate the impact of highly concentrated undiversified asset allocations lies within the scope of robust optimization models. Robust models include methods to improve the accuracy of inputs and to apply robust optimization frameworks to portfolio optimization~\cite{fabozzi2007robust}. Worst-case optimization incorporates uncertainty directly into the optimization process. Along with the development of various robust models, much effort has been devoted to test the performance of these robust portfolios. Although many researchers have conducted out-of-sample performance tests to contrast the classical mean-variance model and robust models, there has not been a dominating conclusion to the performance of such approaches~\cite{kim2013robust}. Our aim in this Thesis is to provide a thorough investigation between the efficiency of classical portfolio
approaches (Value at Risk, Conditional Value at Risk etc) and their robust counterparts.

\section{Contributions}

To the best of our knowledge, no work to date has attacked this specific problem heads on. Thus our main contribution lies in providing an extensive framework for evaluating robust optimization techniques under different architectures. We were particularly interested in investigating the efficiency of the models considered, in periods, which could behave \enquote{out-of-the-box}, by incorporating a high level of turmoil, such as year 2008 where the global financial crisis took place. This addition resulted in an extra degree of complexity, which we should account for in the Thesis, rendering the corresponding results more realistic. At the same time, we compared the performance of the employed robust framework to the non-robust variants of each respective model with respect to certain performance indicators. By doing so, we opt for a concrete understanding in terms of the achieved superiority of robust models as opposed with their non-robust \enquote{enemies}. Overall we can deduce that the performance of the robust models outperforms the non-robust models, for most of the metrics used.

\section{Structure of the Thesis}
\label{sec:structure}

This thesis is organized into the following chapters:
Chapter~\ref{sec:robust} provides a brief overview of robust optimization framework.
Chapter~\ref{sec:portfolio} then describes the portfolio selection problem,
models employed for the computational analysis along with the solution methods( exact and approximate).
Chapter~\ref{sec:results} presents our simulation experiments that we undertake to verify the robustness of the findings; and, finally,
Chapter~\ref{sec:conclusions} concludes and outlines future work.

\chapter{Robust Optimization}
\label{sec:robust}

The field of portfolio selection remains today a particularly active area
of research, which combines elements and methodologies from various fields, such
as optimization, decision analysis under uncertainty, financial risk
management, data science, forecasting etc. Common ground for all these methodologies
used in this specific field is the necessity of modeling and handling the uncertainty in asset returns. To this end, dynamic and stochastic programming models are often employed.~However, an issue that arises with such approaches is that it is often computationally
cumbersome to obtain detailed information about the probability distributions of the
uncertainties in the model. This is a common reason, which explains why dynamic
and stochastic programming methodologies have not become extensively adopted in various applications. Robust optimization is
a recently developed technique, which accounts for the same type of problems as
stochastic programming. However, it typically makes relatively general assumptions on
the probability distributions of the uncertain parameters in order to work with problem
formulations that are more computationally tractable~\cite{fabozzi2007robust}.

Robust optimization extends the framework of traditional linear and non-linear
models (e.g., the mean-variance model), by incorporating instantaneously the
uncertainty as a parameter of the problem. Robust logic deals with
making optimization models robust with respect to constraint violations by solving robust counterparts of these problems for appropriately defined uncertainty
sets for the uncertain parameters. These robust counterparts are in fact worst-case
formulations of the original problem as far as deviations of the parameters from their
nominal values are concerned; however, typically the worst case scenarios are defined
in smart ways that do not lead to overly conservative formulations~\cite{fabozzi2007robust}. An element in
robust optimization is that one makes the problem well defined, by assuming that the
uncertain parameters vary in a particular set defined by one’s knowledge about their
probability distributions and then takes a worst case (max-min) approach. In the optimization literature, the term "robust optimization" has been used to describe several different concepts; however, \textit{robust optimization} refers to an area of optimization whose roots are in the robust control engineering literature.

The way to compute the worst case is also open to debate: should it use a finite number of scenarios, such as historical data, or continuous, convex uncertainty sets, such as polyhedra or ellipsoids? The answers to these questions will determine the formulation and the type of the robust counterpart~\cite{gabrel2014recent}. As pinpointed by Goldfarb and Iyengar~\cite{goldfarb2003robust}, a wide range of robust optimization problems corresponding to the natural class of uncertainty sets (defined by the estimation procedures) can be cast as second-order cone programs (SOCPs) and can be solved very efficiently using interior-point algorithms~\cite{lobo1998applications},~\cite{Nesterov1994},~\cite{sturm1999using}. In fact, both the worst case and practical computational effort required to solve an SOCP is comparable to that for solving a convex quadratic program of similar size and structure; i.e., in practice, the computational effort required to solve these robust portfolio selection problems is comparable to that required to solve the classical Markowitz mean-variance portfolio selection problems~\cite{goldfarb2003robust}.

Our focus in this Thesis is to provide a more recent approach to optimization under uncertainty, in which the uncertainty model is not stochastic, but rather deterministic and set-based~\cite{bertsimas2011theory}.

Given an objective $f_0(\vecx)$ to optimize subject to constraints $f_i(\vecx,\vecu_i) \leq 0$ with uncertain parameters, $\{\vecu_i\}$, the general Robust Optimization formulation is:

\begin{mini}
  {}{f_0(\vecx)}{\label{eq:general}}{}
  \addConstraint{f_i(\vecx,\vecu_i)}{\leq 0}{\quad},  {\forall \vecu_i \in \mathcal{U}_i},{\quad}{i=1,\ldots,m}
\end{mini}

\noindent Here $\vecx \in  \mathbb{R}^N$ is a vector of decision variables, $f_0$, $f_i$: $\mathbb{R}^N \rightarrow \mathbb{R}$ are functions, and the uncertain parameters $\vecu_i \in  \mathbb{R}^k$ are assumed to take arbitrary values in the \textit{uncertainty sets} $\mathcal{U}_i \subseteq \mathbb{R}^k$. The goal of \eqref{eq:general} is to compute minimum cost solutions $\vecx^*$ among all those solutions which are feasible for \textit{all} realizations of the disturbances $\vecu_i$ within~ $\mathcal{U}_i$. Thus, if some of the $\mathcal{U}_i$, are continuous sets, \eqref{eq:general} as stated, has an infinite number of constraints. Intuitively, this problem offers some measure of feasibility protection for optimization problems containing parameters which are not known exactly~\cite{bertsimas2011theory}.\\
Ben-Tal and Nemirovski(~\cite{ben1998robust},~\cite{ben1999robust},~\cite{ben2001lectures}) stated that for suitably defined uncertainty sets $\mathcal{U}$, the robust counterparts of linear programs, quadratic programs, and general convex programs are themselves tractable optimization problems~\cite{goldfarb2003robust}.

\section{Conic Optimization}
\label{sec:conic optimization}

A large set of optimization models, some of which, would be described to a greater extent in the following segments of the Thesis, can be treated as conic optimization problems. A conic optimization (CO) problem (called also conic program) is of the form

\begin{mini}
  {}{\vecc^\top\vecx+d}{}{}
  \addConstraint{\mathbi{A}\vecx}{= \mathbi{b},}{\quad}{\vecx \in \mathbi{K}}
\end{mini}

\noindent where $\vecx \in \mathbb{R}^N $ is the decision vector, $\mathbi{K} \subset \mathbb{R}^M $ represents a closed pointed convex cone with a nonempty interior and the constraint is a given affine mapping from $ \mathbb{R}^n$ to $ \mathbb{R}^m$.\footnote{A set $\mathbi{K}$ is a cone if for all $\vecx \in \mathbi{K}$ it follows that $\alpha\vecx  \in \mathbi{K}$ for all $\alpha\geq0$. A convex cone is a cone with the property that $\vecx + \vecy \in \mathbi{K}$ for all $\vecx , \vecy \in \mathbi{K}$~\cite{fabozzi2007robust}.} Virtually, any convex program can be represented as a conic optimization problem by specifying $\mathbi{K}$ appropriately.

According to Ben-Tal, El Ghaoui and Nemirovski~\cite{ben2009robust}, a wide variety of convex programs are covered by just three types of cones. For the sake of the models presented here, we will narrow down the possible varieties to just one, that being:

\begin{itemize}
\item
Direct products of \textit{Lorentz} (or Second-order, or Ice Cream) cones $\mathbi{L}^k = \{\vecx \in \mathbb{R}^k : \vecx_k \geq \sqrt{\sum_{j=1}^{k-1} \vecx_j^2}\}$. These cones give rise to Conic Quadratic Optimization (called also Second Order Conic Optimization). The mathematical form of a CQO problem is

\begin{mini}
  {}{c^\top\vecx}{}{}
  \addConstraint{\|\mathbi{A}_i\vecx- \mathbi{b}_i\|_2}{\leq \psi_{i}^\top \vecx -d_i,}{\quad}{1\leq i \leq m}
\end{mini}
\end{itemize}

\noindent Keeping in mind the above formulation, we can detect that researchers internationally adopt this form, since it is general enough to encompass linear programs, convex quadratic programs, and constrained convex quadratic programs. At the same time, the problems in this class share many of the properties of linear programs, making the corresponding optimization algorithms used for solving these problems very efficient and highly scalable. Many robust portfolio allocation problems can be formulated as Second Order Cone Programming (SOCP) problems~\cite{fabozzi2007robust}.

\section{Robust Linear Optimization}
\label{sec:robust linear}

A large class of optimization problems can often be cast as robust linear optimization problems by taking the robust counterpart of a linear optimization problem. So, without loss of generality we get:

\begin{mini}
  {}{\vecc^\top\vecx}{\label{eq:rlo}}{}
  \addConstraint{\mathbi{A}\vecx}{\leq \mathbi{b},}{\quad}{\forall \mathbi{a}_1 \in \mathcal{U}_1,\ldots,\mathbi{a}_m \in \mathcal{U}_m},
\end{mini}

\noindent where $\mathbi{a}_i$ represents the $i$-th row of the uncertain matrix $\mathbi{A}$ and takes values in the uncertainty set $\mathcal{U}_i \subseteq \mathbb{R}^N$. Then, $\mathbi{a}_i^\top\vecx\leq\mathbi{b}_i$ ,$\forall \mathbi{a}_i \in \mathcal{U}_i$ if and only if max$_{\{\mathbi{a}_i \in \mathcal{U}_i\}} \mathbi{a}_i^\top\vecx\leq\mathbi{b}_i,~ \forall i$. This is the \textit{subproblem} which must be solved. Ben-Tal and Nemirovski~\cite{ben1999robust} show that the robust LP is essentially always tractable for most practical uncertainty sets of interest. We can't always expect to get a corresponding linear program, by doing this formulation.

\section{Selection of Uncertainty Set}
\label{sec:uncertainty}

A relatively simple way to model uncertainty is to generate scenarios for the possible values of the uncertain parameters using, for example, future asset returns. As explained in Section \ref{sec:robust linear}, scenario optimization can be incorporated in the robust optimization framework by specifying an uncertainty set that is a collection of scenarios for the uncertain parameters.~The robust formulation of the original problem would then contain a set of constraints one of each scenario in the uncertainty set and the optimization would make sure that the original constraint is satisfied for the worst-case scenario in the set~\cite{fabozzi2007robust}. Uncertainty sets are usually extended to richer sets ranging from polytopes to more advanced conic-representable sets derived from statistical procedures. For instance,one can frequently obtain confidence levels for the uncertain parameters~\cite{fabozzi2007robust}. A central feature that Robust Optimization (RO) tries to tackle is the probability guarantees on feasibility under particular distributional assumptions for the disturbance vectors~\cite{bertsimas2011theory}. Specifically, what does robust feasibility imply about the probability of feasibility, i.e. what is the smallest $\epsilon$ we can find such that

\begin{equation*}
\vecx \in X(\mathcal{U}) \Rightarrow \mathbb{P}(f_i(\vecx,\vecu_i)\textgreater 0)\leq \epsilon,
\end{equation*}

\noindent under (ideally mild) assumptions on a distribution for $\vecu_i$?
Such implications may be used as guidance for selection of a parameter representing the size of the uncertainty set.

At this point, we must give the definition introduced by Bertsimas, Brown, and Caramanis~\cite{bertsimas2011theory}, regarding classes of functions $f_i$, coupled with the types of uncertainty sets $\mathcal{U}_i$, that yield tractable robust counterparts. So, the robust feasible set could be

\begin{equation*}
 X(\mathcal{U}) = \{\vecx~|~ f_i(\vecx,\vecu_i) \leq 0, \forall \vecu_i \in \mathcal{U}_i, i = 1,\ldots,m\}.
\end{equation*}	

\subsection{Ellipsoidal Uncertainty}
\label{sec:ellipsoidal}




\textit{Ellipsoidal} uncertainty sets allow for including second moment information about the distributions of uncertain parameters and have been used extensively. Bertsimas, Brown, and Caramanis~\cite{bertsimas2011theory}, mention that controlling the size of these ellipsoidal sets, as in the theorem below, has the interpretation of a budget of uncertainty that the decision-maker selects in order to easily trade-off robustness and performance.

\textbf{THEOREM 2.1} (Ben-Tal and Nemirovski~\cite{ben1999robust}). Let $\mathcal{U}$ be \textit{ellipsoidal} i.e,

\begin{equation*}
\mathcal{U} = U(\Pi,\mathbi{Q}) = \{\Pi(\vecu)| \| \mathbi{Q}\vecu \| \leq \rho\},
\end{equation*}

\noindent \textit{where} $\vecu \rightarrow {\Pi(\vecu)}$ \textit{is an affine embedding of} $\mathbb{R}^L$ \textit{into} $\mathbb{R}^{m \times N}$ \textit{and} $\mathbi{Q} \in \mathbb{R}^{M \times L}$. \textit{Then problem (2.2) is equivalent to an (SOCP). Explicitly, if we have the uncertain optimization}

\begin{mini}
  {}{\vecc^\top\vecx}{}{}
  \addConstraint{\mathbi{a}_i\vecx}{\leq b_i,}{\quad}{\forall \mathbi{a}_i \in \mathcal{U}_i,}{\quad}{\forall i=1\ldots,m},
\end{mini}

\noindent \textit{where the uncertainty set is given as}
\begin{equation*}
\mathcal{U} = \{(\mathbi{a}_1,\ldots,\mathbi{a}_m) : \mathbi{a}_i = \mathbi{a}_i^0 + \Delta_iu_i,  i=1,\ldots,m, \quad \|u\|_2\leq \rho\} 
\end{equation*}

\noindent ($\mathbi{a}_i^0$ \textit{denotes the nominal value), then the robust counterpart is}
\begin{mini}
  {}{\vecc^\top\vecx}{}{}
  \addConstraint{\mathbi{a}_i^0\vecx}{\leq b_i -\rho\|\Delta_i\vecx\|_2,}{\quad}{\forall i=1\ldots,m}.
\end{mini}

\noindent The intuition is the following: for the case of ellipsoidal uncertainty, the subproblem max$_{\{\mathbi{a}_i \in \mathcal{U}_i\}}\mathbi{a}_i^\top\vecx\leq b_i, \forall i$ is an optimization over a quadratic constraint. The dual, therefore, involves quadratic functions, which leads to the resulting SOCP.

\subsection{Polyhedral Uncertainty}
\label{sec:polyhedral}

Polyhedral Uncertainty can be viewed as a special case of ellipsoidal uncertainty~\cite{ben1999robust}. When $\mathcal{U}$ is polyhedral, the subproblem becomes linear, and the robust counterpart is equivalent to a linear optimization problem. We consider the following problem:

\begin{mini}
  {}{\vecc^\top\vecx}{}{}
  \addConstraint{\max_{\mathbi{D}_i\mathbi{a}_i \leq\mathbi{d}_i}\mathbi{a}_i^\top\vecx}{\leq b_i,}{\quad}{ i=1\ldots,m}.
\end{mini}

\noindent The dual of the subproblem ($\vecx$ is not a variable of optimization in the inner max) becomes

\noindent\begin{minipage}{.5\linewidth}
\begin{maxi*}
  {}{\mathbi{a}_i^\top\vecx}{}{}
  \addConstraint{\mathbi{D}_i\mathbi{a}_i}{\leq \mathbi{d}_i}
\end{maxi*}
\end{minipage}%
$\Longrightarrow$
\begin{minipage}{.5\linewidth}
\begin{mini}
  {}{\mathbi{p}_i^\top\mathbi{d}_i}{}{}
   \addConstraint{\mathbi{p}_i^\top\mathbi{D}_i}{= \vecx}
  \addConstraint{\mathbi{p}_i}{\geq 0}
\end{mini}
\end{minipage}\\

\noindent and therefore the robust linear optimization now becomes

\begin{mini}
  {}{\vecc^\top\vecx}{}{}
  \addConstraint{\mathbi{p}_i^\top\mathbi{d}_i}{\leq \mathbi{b}_i,}{\quad i=1, \ldots,m}
    \addConstraint{\mathbi{p}_i^\top\mathbi{D}_i}{= \vecx}{\quad i=1, \ldots,m}
 \addConstraint{\mathbi{p}_i}{\geq 0}{\quad i=1, \ldots,m}.
\end{mini}
   
\noindent Thus the size of such problems grows polynomially in the size of the nominal problem and the dimensions of the uncertainty set.

\subsection{Cardinality Constrained Uncertainty}
\label{sec:cardinality}

A cardinality constraint can be defined as the number of parameters of the problem, which are allowed to deviate from their nominal values. In general, given an uncertainty matrix,  $ \vecA = \begin{pmatrix} 
    a_{ij} 
  \end{pmatrix}$ we assume that each component $a_{ij}$ lies in $[a_{ij} - \hat{a}_{ij}, a_{ij} + \hat{a}_{ij}]$. Following the principle proposed by Bertsimas et al.~\cite{bertsimas2011theory}, we allow at most $\vecG_i$ coefficients of row $i$ to deviate. The positive number $\vecG_i$ controls the trade-off between the optimality of the solution and its robustness to parameter perturbation. Given values $\vecG_1,\ldots,\vecG_m$, the problem is transformed in the robust sense as
  
\begin{mini}
  {}{\vecc^\top\vecx}{}{}
  \addConstraint{\sum_{j}^{}a_{ij}x_j + \max_{S_i\subseteq J_i : |S_i|=\Gamma_i}\sum_{j\in S_i}^{}\hat{a}_{ij}\varrho_j}{\leq \mathbi{b}_i,}{ \quad1\leq i \leq m}
    \addConstraint{-\varrho_j \leq x_j \leq \varrho_j,}{\quad}{\quad1\leq j \leq n}
 \addConstraint{\veclt \leq \vecx \leq\ \veck \quad}{\quad}	
 \addConstraint{\vecrr \geq \mathbf{0} \quad}{\quad}.
\end{mini}

\noindent Because of the set selection in the inner maximization, this problem is nonconvex. We can though, take the dual of the inner maximization problem and acquire an equivalent linear formulation, which is tractable nonetheless. This is what we get:

\begin{maxi}
  {}{\vecc^\top\vecx}{}{}
  \addConstraint{\sum_{j}^{}a_{ij}x_j + \upsilon_i\vecG_i + \sum_{j}^{}n_{ij}}{\leq \mathbi{b}_i,}{ \quad \forall i}
    \addConstraint{\upsilon_i + n_{ij} \geq \hat{a}_{ij}\varrho_j,}{\quad}{\quad \forall i,j}
\addConstraint{-\varrho_j \leq x_j \leq \varrho_j,}{\quad}{\quad \forall j}
\addConstraint{\veclt \leq \vecx \leq\ \veck \quad}{\quad}
\addConstraint{\vecn \geq \mathbf{0} \quad}{\quad}
 \addConstraint{\vecrr \geq \mathbf{0} \quad}{\quad}.
\end{maxi}
  
\subsection{Norm Uncertainty}
\label{sec:norm_uncertainty}

According to Bertsimas, Pachamanova, and Sim~\cite{bertsimas2004robust}, robust linear optimization problems with uncertainty sets described by more general norms can be cast as convex problems with constraints related to the dual norm. Subsequently, we use the notation vec(\vecA) to denote the vector formed by concatenating all of the rows of matrix \vecA~\cite{bertsimas2011theory}.

\textbf{THEOREM 2.2} (Bertsimas, Pachamanova, and Sim~\cite{bertsimas2011theory}). With the uncertainty set 

\begin{equation*}
\mathcal{U} = \left\{ \vecA | ~ \|\vecM(\text{vec}(\vecA) - \text{vec}(\vecAbar))\| \leq \vecD \right\},
\end{equation*}

\noindent where \vecM ~is an invertible matrix, \vecAbar ~is any constant matrix, and $\|\cdot\|$ is any norm, problem \eqref{eq:rlo} is equivalent to

\begin{mini}
  {}{\vecc^\top\vecx}{}{}
  \addConstraint{\vecAbar^\top_i\vecx + \vecD\|(\vecM^\top)^{-1}\vecx_i\|^\ast}{\leq \mathbi{b}_i,}{\quad}{i=1,\ldots,m}
\end{mini}\\

\noindent where $\vecx_i \in \mathbb{R}^{(m\cdot n)\times 1}$ is a vector that contains $\vecx \in \mathbb{R}^n$ in entries $(i-1)\cdot n+1$ through $i\cdot n$ and $0$ everywhere else and $\|\cdot\|^{\ast}$ is the corresponding dual norm of $\|\cdot\|$~\cite{bertsimas2011theory}. Considering the above, Theorem 2.2 leads to an equivalent problem as the norm-based model, with corresponding dual norm constraints. In particular, the $l_1$ and $l_{\infty}$ norms yield linear optimization problems, and the $l_2$ norm results in an SOCP~\cite{bertsimas2011theory}.

\subsection{Perturbation Vectors}
\label{sec:perturbation}

In various applications, a pretty small~(and unavoidable in reality) perturbation of the data may render the nominal optimal solution infeasible~\cite{ben2009robust}. Moreover, a straightforward adjustment of the optimal solution to the actual data may have a negative impact on the quality of the solution. 

For the sake of completeness, we give the definition introduced by Ben-Tal et al.~\cite{ben2009robust}.

$\mathbf{Definition~ 2.1}$ An uncertain Linear Optimization problem is a collection 

\begin{mini}
  {\vecx}{\vecc^\top\vecx + d}{\tag{LOu}\label{perturbation_theory}}{}
 \addConstraint{\mathbi{A}\vecx\leq \mathbi{b}}{}
 \addConstraint{\vecc, d, \mathbi{A}, \mathbi{b} \in \mathcal{U}}{}
\end{mini}

\noindent of LO problems (instances) $\min \left\{\vecc^\top\vecx + d~:\mathbi{A}\vecx\leq \mathbi{b} \right\}$ of common structure (i.e. with common numbers m of constraints and n of variables) with the data varying in a given uncertainty set $\mathcal{U}\subset \mathbb{R}^{(m+1)\times (n+1)}$~\cite{ben2009robust}.

Perturbation vectors are directly associated with the rules of robustness, since they incorporate the uncertainty in a parameterized manner, allowing for some degree of immunization against deviations of the nominal value of the estimated parameter. 

Mathematically, Ben-Tal et al.~\cite{ben2009robust} deal with the uncertainty in an affine fashion, by \textit{perturbation vector} $\Xi$ varying in a given \textit{perturbation set} $\mathcal{J}$ :

\begin{equation}
\renewcommand\arraystretch{1.0}
\mathcal{U} = \left\{
\mleft[
\begin{array}{c|c}
  \vecc^\top & d \\
  \hline
  \mathbi{A} & \mathbi{b}
\end{array}
\mright] = \underbrace{\mleft[
\begin{array}{c|c}
  \vecc_0^\top & d_0 \\
  \hline
  \mathbi{A}_0 & \mathbi{b}_0
\end{array}
\mright]}_{\text{nominal data}\\D_0} + \sum_{l=1}^{L}\Xi_l\underbrace{\mleft[
\begin{array}{c|c}
  \vecc_l^\top & d_l \\
  \hline
  \mathbi{A}_l & \mathbi{b}_l
\end{array}
\mright]}_{\text{basic shifts}\\D_l}~:\Xi \in \mathcal{J}\subset \mathbb{R}_L \right\} 
\end{equation}

\noindent We must note, that the \textit{basic shifts} $D_l$ within the specified uncertainty set $\mathcal{U}$ denote the volatility of the given perturbation vector $\Xi$, within the linear optimization problem. By making this addition to the uncertainty set, we aim to account for unforeseen discrepancies in the performance of the model considered, making the results we get more robust.

Ben-Tal, El Ghaoui and Nemirovski~\cite{ben2009robust} explain that when speaking about perturbation sets with simple geometry \textit{(parallelotopes, ellipsoids)}, we can normalize these sets to be standard. Perturbation vectors are designed for the uncertain data, in a style to generate a reliable solution, which is immunized against uncertainty.

For instance, a parallelotope is by definition an affine image of a unit box $\{\xi \in \mathbb{R}^k :-1\leq\xi_j\leq1, j=1,\ldots,k$ , which allows the opportunity to tackle these problems by using the unit box instead of general parallelotope~\cite{ben2009robust}.
     
\section{Computational Issues}
\label{sec:computational}

Several primal-dual interior-point methods have been developed in the last few years for SOCPs. For instance, Lobo et al.~\cite{lobo1998applications} shows that the number of iterations required to solve a SOCP grows at most as the square root of the problem size, while their practical numerical experiments indicate that the typical number of iterations ranges between 5 and 50 - more or less independent of the problem size~\cite{lobo1998applications}. A feature we want to address in the Thesis is the efficiency of the algorithms emulated for the classical portfolio optimization in comparison with their robust variants. Although robust optimization poses a new trend in portfolio optimization, we want to investigate the complexity of its architecture as opposed to their classic ''enemies''.

\chapter{Portfolio Selection Analysis}
\label{sec:portfolio}

As mentioned in Chapter \ref{sec:robust} the field of portfolio optimization has received much attention among operations researchers. More advanced techniques allow for more complex representations of portfolio procedures under certain conditions at hand. In the present work, we focus on portfolio models, which among others consist of mean-variance framework (MV), which was firstly introduced by the work of  Markowitz~\cite{markowitz1952portfolio}. In its simplest form, mean-variance analysis provides a framework to construct and select portfolios, based on the expected performance of the investments and the risk appetite of the investor. Markowitz reasoned that investors should decide on the basis of a trade-off between risk and expected return. He suggested that risk should be measured by the variance of returns-the average squared deviation around the expected return~\cite{fabozzi2007robust} ,~\cite{markowitz1952portfolio}. Moreover, Markowitz argued that for any given level of expected return, a rational investor would choose the portfolio with minimum variance from the set of all possible portfolios. The set of all possible portfolios that can be constructed is called the \textit{feasible set}. \textit{Minimum variance portfolios} are called \textit{mean-variance efficient portfolios}. The set of all mean-variance efficient portfolios, for different desired levels of expected return, is called the \textit{efficient frontier}. In the following Figure~\ref{portfolio} we provide a graphical depiction of the efficient frontier of risky assets. The feasible set is bounded by the black bold curve.

\begin{figure}[!h]		
  \centering
   \includegraphics[width=1.2\textwidth]{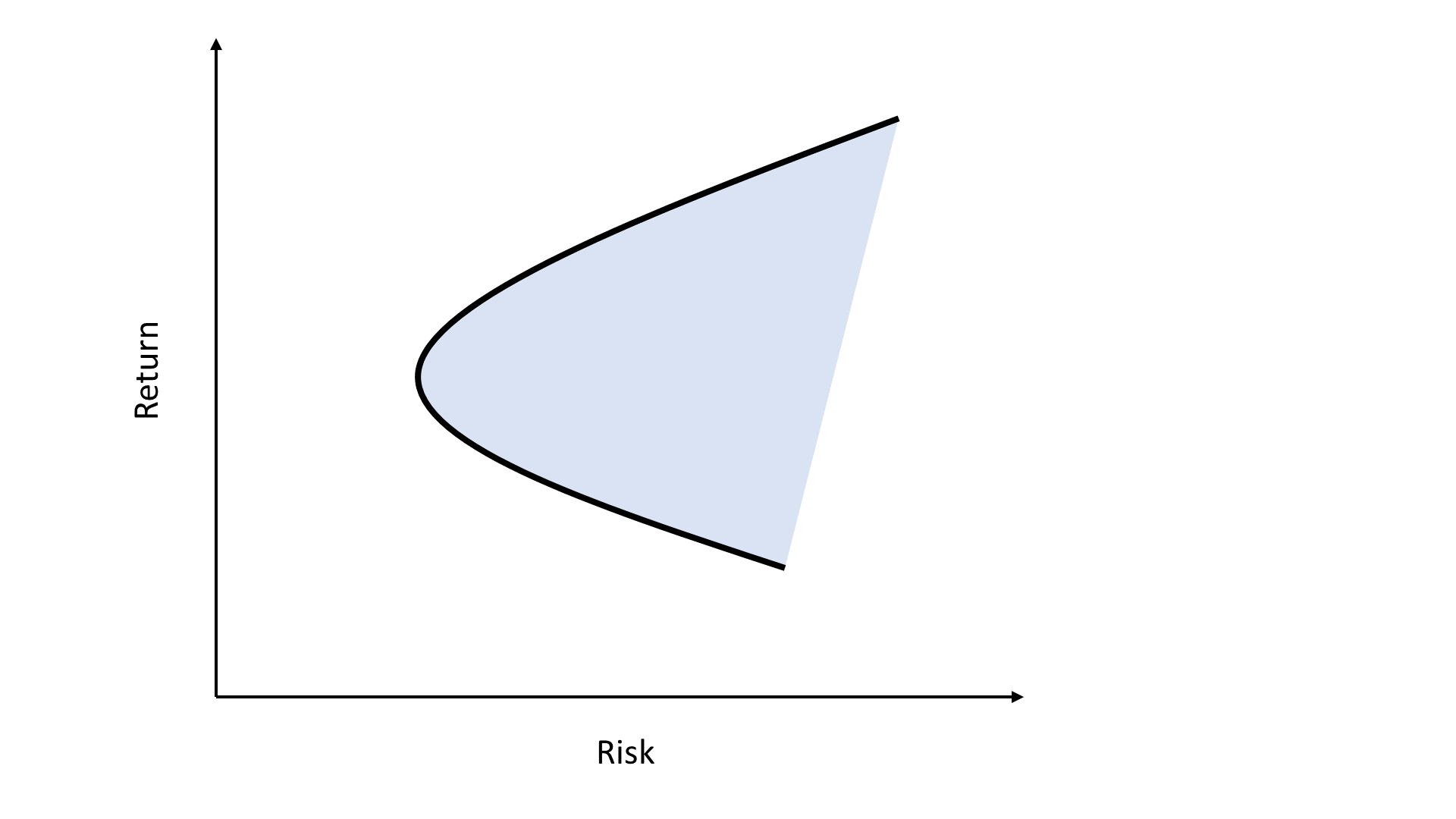}
   \caption{Feasible and Markowitz Efficient Portfolios}
  \label{portfolio}
\end{figure}

\section{Classical Portfolio Optimization }
\label{sec:classical}

Harry Markowitz was the first to model the trade-off between risk and return in portfolio selection as an optimization problem~\cite{markowitz1952portfolio}. However, more than 50 years after Markowitz's seminal work, it appears that full risk-return optimization at the portfolio level is only done at the more quantitative firms, where processes for automated forecast generation and risk control are already in place~\cite{fabozzi2007robust}. From a practical point of view, it is important to make the portfolio selection process robust to different sources of risk-including estimation risk and model risk. Later in the Thesis, we show applications of variations of Markowitz's mean-variance portfolio optimization formulation and show explicitly how to make the problem robust with respect to errors in expected return and covariances estimates. Robustness can be incorporated in the construction of modern portfolios as well, so as to amend their performance in terms of expected returns and the minimization of the respective covariance of returns.

Previously in Section \ref{sec:uncertainty} we gave a brief introduction to uncertainty sets. For the sake of our computational procedure, we incorporate uncertainty about the accuracy of estimates directly in the portfolio optimization process.

\subsection{Mean Variance Optimization}
\label{sec:risk aversion}

In this Section, we give a more detailed explanation of mean-variance models with the underlying mathematical manipulations required to make the corresponding optimization models tractable. First of all, we assume that an investor has to choose a portfolio comprised of $N$ risky assets. Each asset is associated with a respective weight $w_i$, which represents the percentage of the $i$-th asset held in the portfolio and,

\begin{equation*}
\sum_{i=1}^N w_i = 1
\end{equation*}

\noindent We suppose the assets' returns $\mathbi{R}=(R_1,R_2,\ldots,R_N)^\top$ have expected returns \vecm $ = (\textbf{$\mu$}_1,\textbf{$\mu$}_2,\ldots,\textbf{$\mu$}_N)^\top$ and an $N \times N$ covariance matrix given by

\begin{equation*}
\vecS =
\begin{pmatrix}
\sigma_{11} & \sigma_{12} & \ldots
& \sigma_{1N} \\
\sigma_{21} & \sigma_{22} & \ldots
& \sigma_{2N} \\
\vdots & \vdots & \ddots
& \vdots \\
\sigma_{N1} & \sigma_{N2} & \ldots
& \sigma_{NN}
\end{pmatrix}
\end{equation*}

\noindent where $\sigma_{ij}$ denotes the covariance between asset $i$ and asset $j$ such that $\sigma_{ii} = \sigma_{i}^2$, $\sigma_{ij} = \rho_{ij}\sigma_{i}\sigma_{j}$ and $\rho_{ij}$ is the correlation between asset $i$ and $j$. Under these assumptions, the return of a portfolio with weights \vecw $ = (w_1,w_2,\ldots,w_N)^\top \in \mathbf{W} \subseteq\mathbb{R}^N$ is a random variable $R_p  = \vecw^\top\mathbi{R}$ with expected return and variance given by

\begin{equation*}
\mu_p = \vecw^\top\vecm \\
\end{equation*}

\begin{equation*}
\sigma_p^2 = \vecw^\top\vecS\vecw
\end{equation*}

\noindent By picking the portfolio's weights, an investor chooses among the available mean-variance pairs. To calculate the weights for one possible pair, we choose a target mean-return, $\mu_0$. Then, the investor's problem is a constrained minimization problem:

\begin{mini}
  {}{\vecw^\top\vecS\vecw}{}{}
  \addConstraint{\mu_0 = \vecw^\top\vecm}
  \addConstraint{\sum_{i=1}^N w_i = 1}
\end{mini}

\noindent This version of the classical mean-variance optimization problem is known as the \textit{risk minimization formulation}.

However, the mean-variance optimization problem can be expressed in various equivalent forms, in the sense that they all lead to the same efficient frontier as they trade expected portfolio return versus risk in a similar way~\cite{fabozzi2007robust}. Hence, we do present the \textit{Risk Aversion Formulation}.

This alternative formulation explicitly models the trade-off between risk and return in the objective function using a risk aversion coefficient $\lambda$. We denote the following mathematical formulation as (Mv):

\begin{maxi}
  {}{\vecw^\top\vecm - \lambda\vecw^\top\vecS\vecw }{\tag{Mv}\label{first_traditional_model}}{}
  \addConstraint{\sum_{i=1}^N w_i = 1}
\end{maxi}

\noindent When $\lambda$ is small, the penalty from the contribution of the portfolio risk is also small, leading to more risky portfolios. If we gradually increase $\lambda$ from zero and for each instance solve the optimization problem, we end up calculating each portfolio along the efficient frontier. It is a common practice to calibrate $\lambda$ such that a particular portfolio has the desired risk profile. The calibration is often performed via backtests with historical data.

The estimation error in forecasts may significantly influence the resulting optimized portfolio weights.~As explained by Black and Litterman~\cite{black1992global}, small changes in the expected returns, in particular, may have a substantial impact. Indeed, if estimation errors in expected returns are large, they will influence the optimal allocation.

As will be explained later on, mean-variance formulation can be cast with different robust formulations leading to different optimization problems. More specifically,  \textit{Mean Variance with Box Uncertainty} formulation in \ref{sec:uncer_estim}, \textit{Mean Variance with Ellipsoidal Uncertainty} formulation in \ref{sec:counterpart} and \textit{Robust Multi-Objective Optimization} formulation in \ref{sec:multiobjective} are considered variations of classical mean-variance procedure, under different uncertainty set assumptions.

\subsection{Multi-Objective Optimization}
\label{sec:multi}

A major factor we aim to incorporate in the Thesis is the approach established by Fliege and Werner~\cite{fliege2014robust}, where the principle is to start with the multiobjective formulation of the mean-variance portfolio problem. In terms of the robustification procedure, a detailed outline is provided in Section \ref{sec:multiobjective}.

Formally, a multiobjective optimization problem formulation (MOP) is defined as:

\begin{mini}
  {}{\{f_1(\vecw),f_2(\vecw),\ldots,f_k(\vecw)\},\quad \vecw \in \mathbf{W}}{}{} 
\end{mini}

\noindent The feasible set $\mathbf{W}\subseteq \mathbb{R}^N$ is implicitly determined by a set of equality and inequality constraints. The vector function  $\vecf : \mathbb{R}^N \to \mathbb{R}^k $ is composed by $k$ scalar objective functions $f_i : \mathbb{R}^N \to \mathbb{R} (i=1,\ldots,k;k\geq2)$. In multiobjective optimization, the sets $ \mathbb{R}^N $ and  $ \mathbb{R}^k $ are known as decision variable space and objective function space, respectively. The image of $\mathbf{W}$ under the function $\vecf$ is a subset of the objective function space denoted by $\mathcal{Z} = \vecf(\mathbf{W})$ and referred to as the feasible set in the objective function space. In multiobjective optimization problems, there is no canonical order $\mathbb{R}^k$ and thus, we need weaker definitions to compare vectors in $\mathbb{R}^k$~\cite{jaimes2009introduction}. Most of the solution concepts in MCDM come from the old idea of Pareto-efficiency. Any solution is deemed efficient if it is impossible to move to another solution which would improve at least one criterion and make no criterion worse. This can be understood from the following figure.

\pagebreak

\begin{figure}[!h]		
  \centering
   \includegraphics[width=1.1\textwidth]{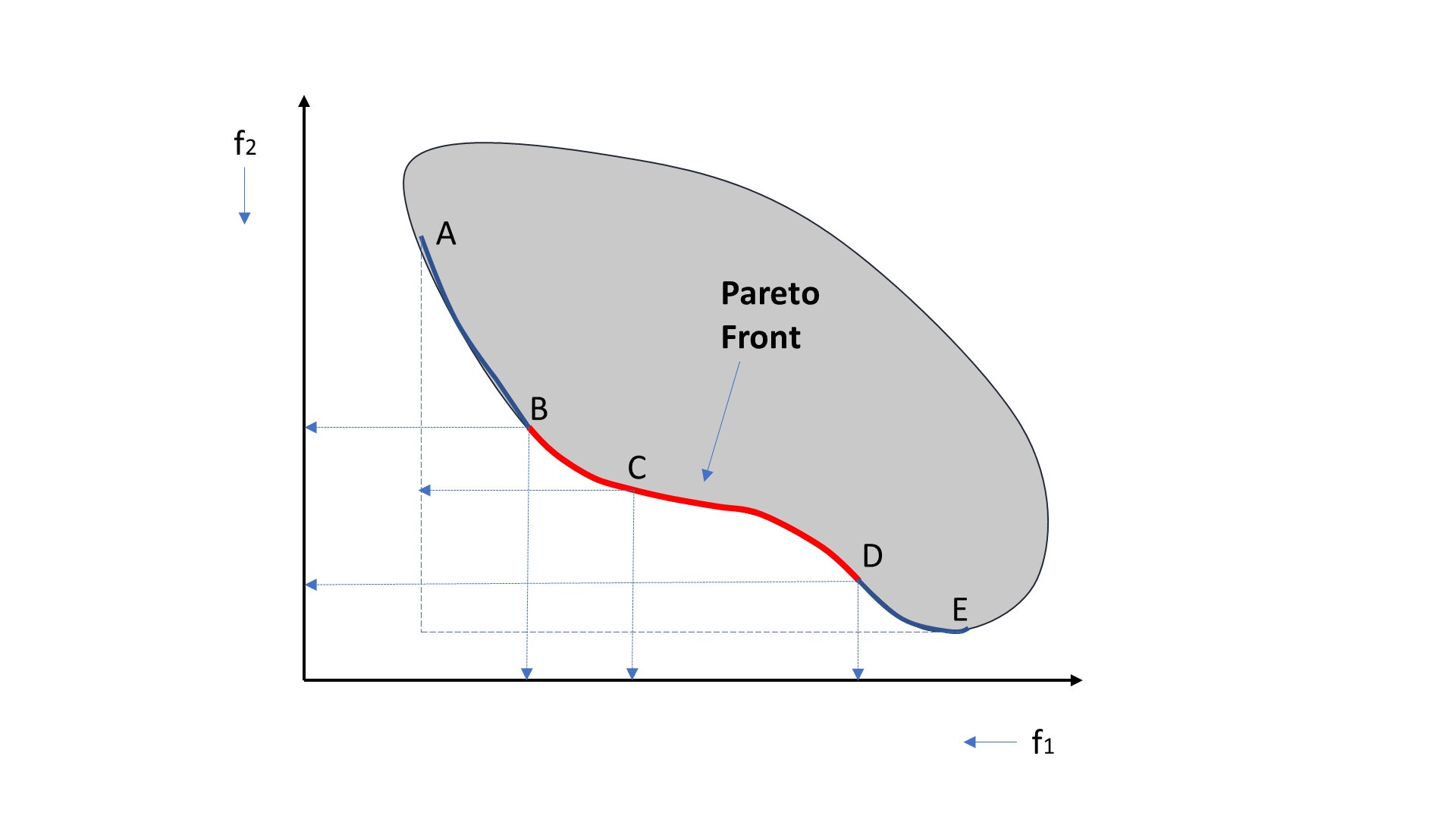}
   \caption{Illustration of the feasible set of solutions}
  \label{pareto}
\end{figure}

Consequently, the feasible set of solutions lies within the Pareto front and are the solutions which will be taken into consideration to designate the optimal pair of solutions from the two objective functions in terms of the quantity we want to optimize at each case. The Pareto front is only composed of non-dominated vectors.

As established by Fliege and Werner~\cite{fliege2014robust}, multicriteria optimization is the ideal setting to analyze portfolio optimization in the sense of Markowitz. From a financial perspective, we can simply set $k=2$, let say $f_1(\vecw) = s(\vecw) = \vecw^\top\vecS \vecw$ be the risk function for some covariance matrix $\vecS$ and let $f_2(\vecw) = -m(\vecw) = -\vecm^\top\vecw$ be the return function for some vector of expected returns $\vecm \in \mathbb{R}^N$.

Fliege and Werner~\cite{fliege2014robust} prove that the generation of almost all efficient portfolios, based on the above assumption can be attained by solving all problems of the form

\begin{mini}
  {}{\lambda_1 s(\vecw) - \lambda_2 m(\vecw) = \lambda_1 \vecw^\top\vecS \vecw - \lambda_2 \vecm^\top \vecw}{\label{eq:multi-scalar}}{} 
  \addConstraint{\vecw \in \mathbf{W}}
\end{mini}

\noindent,  \quad  for all $\lambda_1,\lambda_2 \geq 0$ with $\lambda_1 + \lambda_2 = 1$.

\subsection{Omega ratio Optimization}
\label{sec:omega}

The Omega ratio is a recent performance measure proposed to counterattack the known shortcomings of the Sharpe ratio. The Sharpe ratio is the first attempt to quantify the trade-off between risk and reward in investment under uncertainty~\cite{kapsos2014worst}. However, its underlying assumptions have been widely criticized~\cite{lo2002statistics}. Omega ratio entails the partitioning of the returns into losses and gains in excess of a predetermined threshold and it can be defined as the probability-weighted gains by the probability-weighted losses. In theory, the Omega ratio can be used for any distribution of asset returns. Nevertheless, it assumes accurate knowledge of the distribution. The existence of uncertain input parameters can lead to many solutions for the Omega ratio maximization problem: one solution for each possible realization of the uncertain input. We tackle this problem in the robust sense, by assuming that the realization of the input parameters will be within an uncertainty set. This worst-case approach based on the assumption of only partial information (Ben-Tal and Nemirovski~\cite{ben1998robust}, T\"ut\"unc\"u and Koenig~\cite{tutuncu2004robust}) provides an immunization against the worst-case scenario for all possible realizations of the uncertain input. We establish the worst-case Omega ratio maximization in Section \ref{sec:worst_omega_ratio} under a mixture distribution with uncertain mixing probabilities distributions.

Let $y_i$ denote the random return of asset $i$ and the $i$-th element of the vector $\vecy \in \mathbb{R}^m$. Similarly, $\vecw$ is the vector of weights, whose components add up to 1. The random return of a portfolio of assets is given by $\vecw^\top\vecy$. With $f(y_i)$ and $F(y_i)$, we denote the probability density and cumulative distribution functions, respectively.

According to Keating and Shadwick~\cite{keating2002universal}, Omega ratio is defined as

\begin{equation}
\varOmega(y_i) = \frac{\displaystyle \int_{\tau}^{+\infty}[1-F(y_i)]dy_i}{\displaystyle \int_{-\infty}^{\tau}[F(y_i)]dy_i}
\end{equation}

\noindent The above can be simplified to

\begin{equation}
\label{type_for_Omega}
\varOmega(y_i) = \frac{E(y_i)-\tau}{E[\tau - y_i]^+}+1
\end{equation}

\noindent where $\tau$ denotes a threshold that partitions the returns to desirable (gain) and undesirable (loss). Omega ratio can be distinguished into two categories;~continuous and discrete. In the Thesis, we will deal with the discrete case, where the discrete probability distribution is characterized by a mass function:

\begin{equation*}
\sum_u Pr(\vecy = u) = 1
\end{equation*}

\noindent Considering a discrete probability distribution the Omega ratio for a portfolio is defined as

\begin{equation}
\varOmega(w) = \frac{\vecw^\top(\vecY^\top\vecp)-\tau}{\vecp^\top[\tau\vecones-(\vecY\vecw)]^+}+1
\end{equation}

\noindent where $\vecY \in \mathbb{R}^{S\times N} $ is the matrix that contains the $S$ sample returns for the $N$ assets and $\vecp$ is the vector with the probabilities for each sample return.

As illustrated in Kapsos, Christofides and Rustem~\cite{kapsos2014worst}, the scalar form of the Omega ratio maximization problem employing the linear-fractional programming method under the discrete distribution becomes

\begin{maxi}
  {\vecx,\vecv,\zeta}{\vecx^\top(\vecY^\top\vecp)-\tau \zeta}{\tag{OR}\label{fourth_traditional_model}}{}
   \addConstraint{\vecp^\top\vecv = 1}
   \addConstraint{\vecx^\top(\vecY^\top\vecp)\geq \tau \zeta }  
   \addConstraint{\vecv \geq \tau \zeta\vecones - \vecY\vecx}  
   \addConstraint{\vecv \geq 0} 
   \addConstraint{\sum_{i=1}^N x_i = \zeta} 
   \addConstraint{\zeta\underline{\vecx} \leq \vecx \leq \zeta\xoverline{\vecx}} 
   \addConstraint{\zeta \geq 0} 
\end{maxi}

\noindent Note that the asset weights $\vecw$ have changed to the variables $\vecx$.

\subsection{CVaR Optimization}
\label{sec:cvar}

Since the middle of the 1990s, Value-at-Risk (VaR; RiskMetrics~\cite{morgan1996riskmetrics}), a new measure of downside risk grew in population in financial risk management~\cite{zhu2009worst}. However, several shortcomings were identified in terms of its performance and reliability. Following the conceptual logic behind the mean-variance formulation, here we do present a measure of more complex architecture than VaR, entitled Conditional Value at Risk (CVaR). Defined as the mean of the tail distribution exceeding VaR, CVaR has attracted much attention, since it possesses some better properties than VaR. Let $f(\vecw,\vecy)$ denote the loss associated with the decision vector $\vecw \in \mathbf{W}\subseteq \mathbb{R}^N$ and the random vector $\vecy \in \mathbb{R}^m$. Initially, we assume that $\vecy$ follows a continuous distribution and its corresponding density function is $p(\cdot)$.

Motivated by the theoretical limitations of VaR, Rockafellar and Urysaev~\cite{rockafellar2000optimization} propose an alternative risk measure, CVaR that is defined as the conditional expectation of the loss of the portfolio or equal to VaR \footnote[2]{With more delicate assumptions on portfolio returns, CVaR is also called Extreme Value Theory VaR, Mean-Excess Loss, Mean Shortfall, Tail VaR, Expected Shortfall, or Conditional Tail Expectation (Embrechts et al.~\cite{embrechts2013modelling}; Artzner et al.~\cite{artzner1999coherent})}, that is

\begin{equation}
\label{eq:CVAR}
\text{CVaR}_{\beta}(\vecw) = \frac{1}{1-\text{$\beta$}}{\displaystyle \int_{\small{f(\vecw,\vecy)\geq \text{VaR}_{\beta}(\vecw)}}^{}f(\vecw,\vecy) p(\vecy)dy}
\end{equation}
Rockafellar and Urysaev~\cite{rockafellar2000optimization} prove that CVaR is sub-additive and can be cast for a given confidence level $\beta$ into the following convex optimization problem: $\text{CVaR}_{\beta}(\vecw) = \min_{\alpha\in\mathbb{R}} \mathbf{F}_{\beta}(\vecw,\alpha)$, where $\mathbf{F}_{\beta}(\vecw,\alpha)$ is expressed as 
\begin{equation}
\tag{CVaR}\label{Rocka}
\mathbf{F}_{\beta}(\vecw,\alpha) = \alpha + \frac{1}{1-\text{$\beta$}}{\displaystyle \int_{\small{\vecy\in\mathbb{R}^m }}^{}[f(\vecw,\vecy) - \alpha]^+ p(\vecy)d\vecy}
\end{equation}
where $[\cdot]^+$ is defined as $[t]^+ = \max\{0,t\}$ for any $t\in\mathbb{R}$~\cite{fabozzi2010robust}. 

\noindent Alternatively, CVaR optimization problem can be formulated in a linear programming fashion overriding the integral estimation in \eqref{Rocka}, as pinpointed by Rockafellar and Urysaev~\cite{rockafellar2000optimization}.

\newpage

\begin{mini}
  {}{\xi + \vartheta^{-1}T^{-1}\sum_{i=1}^{T}\nu_{i}}{}{\label{cvar_integral_override}}{}
   \addConstraint{\nu_{i} \geq 0,}{\quad i=1\ldots,T}
  \addConstraint{\nu_{i} \geq f(\vecw,\vecy_{i}) - \xi}{\quad i=1\ldots,T}
\end{mini}

\noindent where $\xi$ is equivalent to $\alpha$ in the initial \eqref{Rocka} form, $\vartheta$ is the same as $1-\text{$\beta$}$ and T denotes the number of the different scenarios.

\noindent Assuming that $f(\vecw,\vecy)$ is linear in $\vecw$, \eqref{cvar_integral_override} is linear and can be solved very efficiently by standard linear programming techniques.

\section{Robust Portfolio Optimization}
\label{sec:robust portfolio}

Having provided a brief review of the traditional portfolio models in Section \ref{sec:classical} our next target is to provide the robust variants of each of the models described above. More specifically, we want to get a more concrete picture of how to construct a portfolio so that the risk is as small as possible with respect to the worst-case scenario of the uncertain parameters. In Chapter \ref{sec:robust} a brief introduction was given in terms of tractable reformulations of the uncertainty. Here, we survey some recent advances in portfolio selection with parameter uncertainty. 

\subsection{Mean Variance with Box Uncertainty}
\label{sec:uncer_estim}

A reasonable way to incorporate uncertainty caused by estimation discrepancies is to require that the investor be protected if the estimated return $\hat{\mu_i}$ for each asset is around the true expected return ${\mu_i}$. The error from the estimation can be assumed to be not larger than some small number $\delta_i \geq 0$. The simplest possible choice for the uncertainty set for $\vecm$ is the \enquote{box}~\cite{fabozzi2007robust}.

\begin{equation*}
\mathcal{U}_{\delta}(\vecmhat) = \{\vecm|\mu_i - \hat{\mu_i}|\leq \delta_i ,i=1,\ldots,N\}
\end{equation*}

\noindent A rational treatment for $\delta_i's$ could be the incorporation of some confidence level around the estimated expected return. In our case, we consider that the individual return of the risky assets is normally distributed, meaning that $\frac{\mu_i - \hat{\mu_i}}{\sigma_{i}/\sqrt{T}_i}$ follows a standard normal distribution, and a $95$\% confidence level for ${\mu_i}$ can be obtained by setting $\delta_i = 1.96\sigma_{i}/\sqrt{T}_i$, where $T_i$ is the sample size used in the estimation and $\sigma_{i}$ is the standard deviation of asset $i$. Subsequently, we do get the robust formulation of the mean-variance problem under the assumption on ${\mu_i}$ , which we denote with (MvBU).

\begin{maxi}
  {}{\vecw^\top\vecmhat - \vecd^\top|\vecw| - \lambda\vecw^\top\vecS\vecw }{\tag{MvBU}\label{first_robust_model}}{}
    \addConstraint{\sum_{i=1}^Nw_i = 1}   
\end{maxi}

\noindent As explained in~\cite{fabozzi2007robust}, if the weight of asset $i$ in the portfolio is negative, the worst-case expected return for asset $i$ is $\mu_i + \delta_i$ (we lose the largest amount possible). If the weight of asset $i$ in the portfolio is positive, then the worst-case expected return for asset $i$ is $\mu_i - \delta_i$ (we gain the smallest amount possible). In this robust version of the mean-variance formulation, assets whose mean return estimates are less accurate (have a larger estimation error $\delta_i$) are penalized in the objective function and would prospectively lead to having smaller weights in the optimal portfolio allocation.

\subsection{Mean-Variance with Ellipsoidal Uncertainty}
\label{sec:counterpart}

Even though more general uncertainty sets lead to more complicated optimization problems, the intuition behind the uncertainty remains the same.~We introduce an alternative form of the uncertainty which is incorporated in the expected returns vector $\vecm$, the \textit{ellipsoidal uncertainty}.

\begin{equation*}
\mathcal{U}_{\delta}(\vecmhat) = \Big\{\vecm|(\vecm - \vecmhat)^\top\vecS_{\mu}^{-1}(\vecm - \vecmhat)\leq\delta^2\Big\}
\end{equation*}

\noindent where $\vecmhat$ is the vector of mean estimated returns,  $\vecm$ is the vector of mean true returns from all the stocks considered respectively and $\vecS_{\mu}$ represents the covariance matrix of the errors in the estimation of the \textit{expected} (average) returns.

The adoption of this specific uncertainty set envisages the idea that the investor would like to be protected in instances in which the total scaled deviation of the realized average returns from the estimated returns is within $\delta$. This uncertainty set cannot be interpreted as individual confidence levels around each point estimate. However, its representation resembles a joint confidence region used, for example, in Wald tests~\cite{Fabozzi2007}. For the emulated model we approximate $\delta$ with the inverse $\large{\rchi^2}$ distribution. Similarly to the first specification of uncertainty (box), here again, we want to detect the "worst" estimates of the expected returns and how would this impact the allocation of the portfolio. Mathematically, this can be expressed as

\begin{maxi}
  {w}{\min_{\small{\vecm} \in \Big\{\vecm|(\vecm - \vecmhat)^\top\vecS_{\mu}^{-1}(\vecm - \vecmhat) \leq  \mathlarger \delta^2\Big\}}\vecw^\top\vecmhat - \lambda\vecw^\top\vecS\vecw }{\tag{MvEU}\label{second_robust_model}}{}
   \addConstraint{\sum_{i=1}^Nw_i = 1}     
\end{maxi}

\noindent We denote this problem as (MvEU), which stands for the Mean Variance with Ellipsoidal Uncertainty formulation and is not in a form that can be input into a standard optimization solver. We need to solve the "inner" problem first while holding the vector of weight $\vecw$ fixed and compute the worst expected portfolio return over the set of possible values for $\vecm$.

The robust problem that occurs after some algebra manipulations is the following:

\begin{maxi}
  {w}{\vecw^\top\vecm - \lambda\vecw^\top\vecS\vecw - \delta\sqrt{\vecw^\top\vecS_{\mu}\vecw}}{}{}
   \addConstraint{\sum_{i=1}^Nw_i = 1}  
\end{maxi}

\noindent Just as in the previous problem, we interpret the term $\delta\sqrt{\vecw^\top\vecS_{\mu}\vecw}$ as the penalty for estimation risk, where $\delta$ reflects the degree of the investor's aversion to estimation risk. It is not immediately obvious how one can estimate $\vecS_{\mu}$. According to Fabozzi, Kolm, Pachamanova and Focardi~\cite{Fabozzi2007} critics of this approach have argued that the realized returns typically have large stochastic components that belittle the expected returns, and hence estimating $\vecS_{\mu}$ accurately from historical data is very hard, if not impossible~\cite{lee2006robust}. Several approximate methods for estimating $\vecS_{\mu}$ have been found to work well in practice~\cite{stubbs2005computing}. In our case, assuming that returns in a given sample of size T come from a normal distribution, we consider $\vecS_{\mu} = (1/T)\cdot\vecS$ ~\cite{Fabozzi2007}, where $\vecS$ is the covariance matrix of asset returns as depicted in Section \ref{sec:risk aversion}.

At this phase, we provide the Cholesky Decomposition of the covariance matrix $\vecS_{\mu}$ for making the optimization problem above solver-friendly and thus converting it into a Second Order Cone Programming (SOCP) problem. We introduce a new variable $\vecz$ to replace the quantity $\delta\sqrt{\vecw^\top\vecS_{\mu}\vecw}$ and another one variable $\vecq$. As occurs from this transformation of the problem at hand, $\vecS_{\mu} = \vecC^\top\vecC$. After these reformulations, the Cholesky Decomposition for this problem can be expressed as

\begin{maxi}
  {}{\vecw^\top\vecm - \lambda\vecw^\top\vecS\vecw - \delta\vecz}{}{}
   \addConstraint{\vecC\vecw - \vecq = 0}
   \addConstraint{\sum_{i=1}^Nq_i^2 - \vecz^2 \leq 0}  
   \addConstraint{\sum_{i=1}^Nw_i = 1}  
   \addConstraint{\vecw\geq0,\vecq,\vecz \in \mathbb{R}} 
\end{maxi}

\subsection{Robust Multi-Objective Optimization}
\label{sec:multiobjective}

The general setting we employ for this class of problems is the following convex parametric optimization problem

\begin{align*}\label{robustification_mult}
  &\underset{\small{\vecw} \in \mathbf{W}}{\text{efmin}} \quad f(\vecw,u)\\
  &\text{s.t.} \quad \quad  g(\vecw,u) \leq 0
\end{align*}

\noindent In the above formulation, it is assumed that $\mathbf{W}$ subsumes all certain constraints, whereas all uncertain constraints explicitly depending on u are handled by the inequality $g(\vecw,u) \leq 0$~\cite{fliege2014robust}. The operator efmin in a given non-empty and convex set M $\subset \mathbb{R}^N$ is searching for efficient points at each order relation according to the dimensions considered in the application at hand, as explained by Fliege and Vicente~\cite{fliege2006multicriteria} .
 
Equivalently, the \textit{multi-objective robust counterpart} for the multi-objective problem with uncertainties (as formulated above) is defined as 

\begin{align*}
  &\underset{\small{\vecw} \in \mathbf{W}}{\text{efmin}} \quad f_U^{RC}(\vecw)\\
  &\text{s.t.} \quad \quad g_U^{RC}(\vecw) \leq 0
\end{align*}

\noindent with

\noindent
\makebox[\linewidth]{%
  \begin{minipage}[t]{\dimexpr0.5\linewidth-.2pt}
    \vspace{-\baselineskip}
    \begin{align*}
      f_U^{RC}(\vecw) :=
\begin{pmatrix}
\max_{u \in U}f_1(\vecw,u) \\
\vdots \\
\max_{u \in U}f_k(\vecw,u) \\
\end{pmatrix}
    \end{align*}
  \end{minipage}%
  \vrule
  \begin{minipage}[t]{\dimexpr0.5\linewidth-.2pt}
    \vspace{-\baselineskip}
    \begin{align*}
      g_U^{RC}(\vecw) :=
\begin{pmatrix}
\max_{u \in U}g_1(\vecw,u) \\
\vdots \\
\max_{u \in U}g_m(\vecw,u) \\
\end{pmatrix}
    \end{align*}
  \end{minipage}
}

\vspace{\belowdisplayskip}

\noindent i.e. each component of the objective function and the constraints is replaced by its robust counterpart. We realize that for $k=1$ the definition of the multiobjective robust counterpart coincides with the definition of the usual robust counterpart and hence provides a proper generalization of this concept to multiobjective optimization. As pinpointed by Fliege~\cite{fliege2014robust} the robustification of the $k$-dimensional objective is now the very same as the robustification of the $m$-dimensional constraint.

Here we do consider the formulation \eqref{eq:multi-scalar} from the perspective of uncertainty. There have been many studies focusing on the uncertainty for the expected returns alone, others on covariance alone by specifying confidence levels with lower and upper bounds for individual elements. In this work, we will consider an uncertainty set, which will include expected returns and covariance in a unified way. No estimated parameter acts independently and the task is to examine their efficiency when these two estimated parameters deviate from their real value. For simplicity of the exposition that follows, we choose an ellipsoid around a nominal point $(\vecmhat,\vecShat)$ of size $\varepsilon$. 

\begin{equation}\label{robust_mult}
  \mathcal{U}_{\varepsilon}(\vecmhat,\vecShat)=\{(\vecm,\vecS) \in \mathbb{R}^N \times \mathbb{S}^N_{+} : \|\vecm - \vecmhat\|+c\|\vecS-\vecShat\|\leq\varepsilon\}
\end{equation}

\noindent where $\mathbb{S}^N_{+}$ denotes the cones of positive semidefinite matrices.\\

\noindent Our main motivation for the consideration of multiobjective problems under data uncertainty stems from mean-variance portfolio optimization.~In this present work, what will be of our primary importance, is the reformulation of a multiobjective model into its robust variant, which will incorporate the uncertainty as stated in equation \eqref{robust_mult}. The model we will consider is adopted from Fliege and Werner ~\cite{fliege2014robust}.

Taking into account the uncertainty in expected returns and in the covariance matrix, we use the previously introduced joint uncertainty set as shown in \eqref{robust_mult} and derive the following robust multiobjective mean-variance formulation:

\begin{mini*}
{}{\vecw^\top\vecS\vecw + \frac{\varepsilon}{c} \left\| \vecw \right\|^{2}}
{\label{eq:first}}
{f_1 = }
\end{mini*}
\begin{mini}
{}{-\vecm^\top\vecw + \varepsilon\left\| \vecw \right\|}
{\label{eq:second}}
{f_2 = }
\addConstraint{\sum_{i=1}^Nw_i = 1}  
\addConstraint{\vecw \geq 0}
\end{mini}
	
\noindent Based on the specific choice of the uncertainty set \eqref{robust_mult} the robustified versions (i.e. the robust counterparts) of $s(\vecw) = \vecw^\top\vecS \vecw$ and $-m(\vecw) = -\vecm^\top\vecw$ can be analytically obtained as :

\begin{align}
\label{eq:tcheb}
\begin{split}
s^{RC}(\vecw) &= \max\limits_{\left(\vecm,\vecS\right) \mathlarger\in \mathcal{\mathlarger U}_{\mathlarger\varepsilon}\left(\vecmhat,\vecShat\right)}\vecw^\top\vecS\vecw = \vecw^\top\left(\vecShat + \frac{\varepsilon}{c}I\right)\vecw = \vecw^\top\vecShat\vecw + \frac{\varepsilon}{c}\|\vecw\|^2\\
 -m^{RC}(\vecw) &= \max\limits_{\left(\vecm,\vecS\right) \mathlarger\in \mathcal{\mathlarger U}_{\mathlarger\varepsilon}\left(\vecmhat,\vecShat\right)}-\vecm^\top\vecw = -\vecmhat^\top\vecw + \varepsilon\|\vecw\|
 \end{split}
\end{align}	

\subsubsection{Chebyshev Scalar Transformation}
\label{sec:tchebycheff}

We employ the Chebyshev scalarizing function proposed by Steuer and Choo~\cite{steuer1983interactive} in order to render \eqref{eq:second} in a structure that could be efficiently solved by a standard optimization software. For the robust counterpart problem in \eqref{eq:tcheb} we have two objective functions that we aim to maximize. So, respectively $\eta_1,\eta_2 \geq 0$ are given weights such that $\eta_1+\eta_2=1$. The outline of the scalarization technique applied in \eqref{eq:tcheb} is the following :

\begin{mini}
{}{\alpha}
{\label{eq:tcheb_sec}}
{}
\addConstraint{\alpha \geq \eta_1 \Big\{\vecw^\top\left(\vecS + \frac{\varepsilon}{c}I \right)\vecw - f_1^\ast \Big\}}
\addConstraint{\alpha \geq \eta_2 \Big\{f_2^\ast - (\vecm^\top\vecw -\varepsilon\|\vecw\|) \Big\}}
\addConstraint{\sum_{i=1}^Nw_i = 1} 
\addConstraint{\vecw,\alpha \geq 0}
\end{mini}

\noindent where $f_{1}^{\ast}$ and $f_{2}^{\ast}$ are the optimal values of the problems below:

\noindent\begin{minipage}{.5\linewidth}
\begin{mini*}
  {}{\vecw^\top\left(\vecS + \frac{\varepsilon}{c}I \right)\vecw}{}{f_1^\ast =}
  \addConstraint{\sum_{i=1}^Nw_i = 1} 
  \addConstraint{\vecw \geq 0}
\end{mini*}
\end{minipage}%
\begin{minipage}{.5\linewidth}
\begin{maxi}
  {}{\vecm^\top\vecw - \varepsilon\|\vecw\|}{}{f_2^\ast =}
  \addConstraint{\sum_{i=1}^Nw_i = 1} 
  \addConstraint{\vecw \geq 0}
\end{maxi}
\end{minipage}\\

\noindent In model \eqref{eq:tcheb_sec} $\eta_1$ and $\eta_2$ represent the the weights of the two objective functions, accordingly. This is a convex non-linear problem and can't be fed to the software in this form. By adding a  pseudo-parameter $\omega$, we treat this problem into a more comprehensive mathematical form, which we denote as RMu :

\begin{mini}
{}{\alpha}
{\tag{RMu}\protect\label{third_robust_model}}
{}
\addConstraint{\Big\{\vecw^\top\left(\vecS + \frac{\varepsilon}{c}I \right)\vecw - \frac{1}{\eta_1}\alpha \leq f_1^\ast \Big\}}
\addConstraint{\Big\{\vecm^\top\vecw + \frac{1}{\eta_2}\alpha- \delta\omega = f_2^\ast \Big\}}
\addConstraint{\|\vecw\|^2 - \omega^2 \leq 0}
\addConstraint{\sum_{i=1}^Nw_i = 1} 
\addConstraint{\vecw,\alpha,\omega \geq 0}
\end{mini}

\subsection{Worst-case Omega ratio}
\label{sec:worst_omega_ratio}

As explained in Section \ref{sec:omega} the optimization of Omega ratio requires exact knowledge of the probability distribution of asset returns $\vecy$. Since partial knowledge of estimation errors can lead to overoptimistic solutions, we introduce the worst-case Omega ratio.

The worst-case Omega ratio (WO) for a fixed $\vecw \in \mathbf{W}$ with the assumption of the discrete analog of a set of probability distributions is defined as~\cite{kapsos2014worst}

\begin{equation}
\text{WO}(\vecw) \equiv \inf_{\small{\vecp} \in \Pi} \frac{\vecw^\top(\vecY^\top\vecp)-\tau}{\vecp^\top[\tau\vecones-(\vecY\vecw)]^+}
\end{equation}

\noindent where the density function is only known to belong to a set $\Pi$ of distributions.

We do consider the mixture distribution uncertainty, where it is known that the underlying distribution is a mixture distribution with known continuous mixture components but unknown mixture weights. We employ the efficient frontier approach. Mixture distribution is defined as a convex combination of probability density functions, known as mixture components. The weights associated with the mixture components are called mixture weights. First of all, we assume that the distribution of $\vecy$ is characterized by the mixture of a set of prespecified distributions with unknown mixture weights. So,

\begin{equation*}
\varLambda \equiv \Big\{\vecl = (\lambda_1,\ldots,\lambda_l): \sum_{i=1}^l\lambda_i = 1, \lambda \geq 0, i=1,\ldots,l\Big\}.
\end{equation*}

\noindent Let the distribution of $\vecy$ being characterized by a mixture of a set of distributions with unknown mixing parameters such that

\begin{equation*}
p(\vecy) \in \mathcal{P} = \Big\{\sum_{i=1}^l\lambda_ip^i(\vecy):\lambda \in \varLambda\Big\}
\end{equation*}

\noindent where $\lambda_i$ is the unknown mixture weight of the probability distribution $p^i(\vecy)$.

\noindent Employing a robust counterpart approach and using the efficient frontier method~\cite{kapsos2014optimizing} the optimization program becomes:

\begin{maxi}
{\vecw \in \mathbf{W},\theta \in \mathbb{R}}{\theta}{\protect\label{eq:worst_omega_1}}{}
\addConstraint{\gamma(\vecw^\top~E_p(\vecy)^i-\tau) - (1-\gamma)E_{p^i}([\tau - \vecw^\top~\vecy]^+) \geq \theta,}{\quad}{\forall i=1,\ldots,l}
\end{maxi}

\noindent In order to obtain the portfolio with the maximum worst-case Omega ratio, the above problem needs to be solved for different values of $\gamma$. An algorithm for performing this task is presented below.\\

\begin{algorithm}[!h]
 \caption{Designation of maximum worst-case Omega ratio~\cite{kapsos2014worst}}
  	Set $\gamma = 0$, wcor $ = -\infty, \vecw^{\ast} = 0$ \\
	 \While {$\gamma \leq 1$}{
		Solve \eqref{eq:worst_omega_1} and get $\vecw_{candidate}$\\
		Set	minOR $=$ min\{Omega ratio for each distribution\}\\
		 \If {minOR $>$ wcor}{
		 	wcor $=$ minOR, $\vecw^{\ast}$ $=$ $\vecw_{candidate}$
		 }
		 $\gamma = \gamma + step$
	}
	\Return $\vecw^{\ast}$, wcor
\end{algorithm} 

\noindent Different uncertainty sets may lead to significantly different decisions. The trade-off between robustness and performance must be taken into account. Under mixture distribution uncertainty, the modeler has to determine the mixture components. In this current work, the results obtained are contingent on the analysis of historical data using different subsets.

We do present the worst-case Omega ratio application under mixture distribution, which will be applied to real data as shown more explicitly in Chapter \ref{sec:results}. We are employing the discrete analog of \eqref{eq:worst_omega_1}, as described in Kapsos et al.~\cite{kapsos2014optimizing}, i.e.

\begin{maxi}
{}{\theta}
{\tag{WCOR}\protect\label{fourth_robust_model}}{}
\addConstraint{\gamma(\vecw^\top\vecm^i-\tau) - (1-\gamma)\frac{1}{S^i}\vecones^\top\vecu^i \geq \theta, \forall i=1,\ldots,l}
\addConstraint{\vecu^i \geq \tau\vecones - \vecY^i\vecw, \forall i=1,\ldots,l}
\addConstraint{\vecu^i \geq 0 , \forall i=1,\ldots,l}
\addConstraint{0 \leq \vecw \leq 1}
\addConstraint{\sum_{i=1}^Nw_i = 1}
\end{maxi}

\noindent where $\vecm^i$ is the vector with the expected returns for the $i$-th mixture component, $S^i$ is the number of samples from the $i$-th mixture component, $\vecu^i$ an auxiliary variable introduced to linearize the max function in \eqref{fourth_robust_model} and $\vecY^i$ the $S^i\times N$ matrix that contains the sample returns from the $i$-th distribution for the $N$ assets.

\subsection{Worst-case CVaR}
\label{sec:worst_cvar}

In this Section, we assume that the density function of the portfolio return $p(\cdot)$ is only known to belong to a certain set $\mathcal{P}$ of distributions, i.e., $p(\cdot)\in\mathcal{P}$. Zhu and Fukushima~\cite{zhu2009worst} define the worst-case CVaR (WCVaR) for fixed $\vecw\in\mathbf{W}$ with respect to $\mathcal{P}$ as:

\begin{equation*}
\text{WCVaR}_{\beta}(\vecw) = \sup_{p(\cdot)\in\mathcal{P}}\text{CVaR}_{\beta}(\vecw)
\end{equation*}

\noindent where the computation of CVaR was explained in closer detail in Section \ref{sec:cvar}.

\subsubsection{Mixture Distribution}
\label{sec:mixture}

As explained previously for the Worst-case Robust Omega ratio in Section \eqref{sec:worst_omega_ratio}, here we do assume that the density function of $\vecy$ is only known to belong to a set of distributions which consists of all the mixture distributions of some predetermined likelihood distributions, i.e.,
 
\begin{equation}
\label{eq:mixture_set}
p(\cdot)\in\mathcal{P} \overset{\mathrm{\Delta}}{=} \Big\{\sum_{i=1}^l\lambda_ip^i(\cdot):\sum_{i=1}^{l}\lambda_i = 1, \lambda_i \geq 0, i=1,\ldots,l\Big\}
\end{equation}

\noindent where $p^i(\cdot)$ signifies the $i$-th distribution scenario, and $i$ denotes the number of possible scenarios.

\noindent With respect to the uncertainty set we have defined, we realize that

\begin{equation}
\label{eq:worst_cvar}
\mathbf{F}_{\beta}^i(\vecw,\alpha) = \alpha + \frac{1}{1-\text{$\beta$}}{\displaystyle \int_{\small{\vecy\in\mathbb{R}^m }}^{}[f(\vecw,\vecy) - \alpha]^+ p^i(\vecy)d\vecy}. \quad i=1,\dots,l
\end{equation}

\noindent We reformulate the original problem to a more tractable one. It can be seen that the WCVaR minimization is equivalent to

\begin{mini}
{\vecw,\alpha,\theta}{\theta}{\protect\label{eq:worst_case_cvar}}{}
\addConstraint{\alpha + \frac{1}{1-\text{$\beta$}}\displaystyle \int_{\vecy\in \mathbb{R}^m}{}[f(\vecw,\vecy) - \alpha]^+p^i(\vecy)d\vecy \leq \theta,}{\quad}{i=1,\ldots,l}
\end{mini}

\noindent An approximation method can be used to tackle the difficulty of the computation of the integral of a multivariate and nonsmooth function in \eqref{eq:worst_case_cvar}. Zhu et al.~\cite{zhu2009worst} mention that Monte Carlo simulation is one of the most effective methods for high-dimensional integral calculation. Rockafellar and Urysaev~\cite{rockafellar2000optimization} use this method to approximate $\mathbf{F}_{\beta}(\vecw,\alpha)$ as

\begin{equation}
\label{eq:sampling}
\vecFtilde_{\beta}(\vecw,\alpha) = \alpha + \frac{1}{S(1-\text{$\beta$})}\displaystyle \sum_{k=1}^{S}[f(\vecw,\vecy_{[k]}) - \alpha]^+
\end{equation}

\noindent where $\vecy_{[k]}$ is the $k$-th sample generated by simple random sampling with respect to $\vecy$ according to its density function $p(\cdot)$, and $S$ denotes the number of samples. 

\noindent Replacing the integral in \eqref{eq:worst_case_cvar} with \eqref{eq:sampling} yields

\begin{mini}
{\vecw,\alpha,\theta}{\theta}{\protect\label{eq:sampling_res}}{}
\addConstraint{\alpha + \frac{1}{S^i(1-\text{$\beta$})}\sum_{k=1}^{S^i}[f(\vecw,\vecy_{[k]}^i) - \alpha]^+ \leq \theta,}{\quad}{i=1,\ldots,l}
\end{mini}

\noindent where $\vecy_{k}^i$ denotes the $k$-th sample with respect to the $i$-th distribution scenario $p^i(\cdot)$ and $S^i$ denotes the number of corresponding samples. The approximation of problem \eqref{eq:worst_case_cvar} could as well be formulated as 

\begin{mini}
{\vecw,\alpha,\theta}{\theta}{\protect\label{eq:worst_case_cvar_approxi}}{}
\addConstraint{\alpha + \frac{1}{1-\text{$\beta$}}\sum_{k=1}^{S^i}\pi_{k}^i[f(\vecw,\vecy_{[k]}^i) - \alpha]^+ \leq \theta,}{\quad}{i=1,\ldots,l}
\end{mini}

\noindent where $\pi_{k}^i$ denotes the probability according to the $k$-th sample with respect to the $i$-th likelihood distribution $p(\cdot)^i$. If $\pi_{k}^i$ is equal to $\frac{1}{S^i}$ for all $k$, then \eqref{eq:worst_case_cvar_approxi} reduces to \eqref{eq:sampling_res}. We denote $\vecp^i = (\pi_{1}^i,\ldots,\pi_{S^i}^i)^\top$.

\noindent By introducing some new variables the optimization problem \eqref{eq:worst_case_cvar_approxi} can be rewritten as the following minimization problem with variables $(\vecw,\vecu,\alpha,\theta) \in \mathbb{R}^N\times \mathbb{R}^m\times \mathbb{R}\times \mathbb{R}$.

\begin{mini}
{}{\theta}{\protect\label{eq:worst_case_cvar_next}}{}
\addConstraint{\alpha + \frac{1}{(1-\text{$\beta$})}(\vecp^i)^\top\vecu^i}{\leq \theta,}{\quad}{i=1,\ldots,l}
\addConstraint{\vecu_{k}^i \geq f(\vecw,\vecy_{[k]}^i) - \alpha,}{\quad}{\quad}  {k=1,\ldots,S^i, i=1\ldots,l}
\addConstraint{\vecu_{k}^i\geq 0,}{\quad}{\quad}{k=1,\ldots,S^i, i=1\ldots,l}
\addConstraint{\vecw \in \mathbf{W}}{\quad}
\end{mini}

\noindent As defined in Section \ref{sec:omega} the random vector $\vecy = (y_1,y_2,\ldots,y_N)^\top\in\mathbb{R}^N$ represents the uncertain returns of the N risky assets. Adjusting the above formulation to incorporate some additional elements, we consider the loss function to be defined as

$f(\vecw,\vecy) = -\vecw^\top\vecy$

\noindent Portfolio optimization tries to locate an optimal trade-off between the risk and the return according to the investor's preference, whereas the robust portfolio selection is performed through the worst-case analysis of risk and return~\cite{zhu2009worst}. Thus, the robust portfolio selection problem using WCVaR as a risk measure can be represented as 

\begin{equation*}
\min_{\small{\vecw\in\mathbf{W}}}\text{WCVaR}(\vecw)
\end{equation*}

\noindent We complete the formulation of robust portfolio selection model, by specifying the constraint set $\mathbf{W}$. We suppose that the investor has an initial wealth $w_0$. Thus the portfolio selection satisfies

\begin{equation*}
\vece^\top\vecw = w_0
\end{equation*}

\noindent In the case of mixture distribution uncertainty given by \eqref{eq:mixture_set} let $\vecytilde^i$ denote the expected value of $\vecy$ with respect to the likelihood distribution $p(\cdot)$. In terms of the worst-case minimum expected return $\phi$ required by the investor the following condition must hold

\begin{equation}
\vecw^\top\vecytilde^i\geq \phi, i=1,\ldots,l
\end{equation}

\noindent The robust portfolio selection problem, under the mixture distribution probability, is formulated as the following linear program with variables $(\vecw,\vecu,\alpha,\theta) \in \mathbb{R}^N\times \mathbb{R}^m\times \mathbb{R}\times \mathbb{R}$.\\

\begin{mini}
{}{\theta}
{\tag{WCVaR}\protect\label{fifth_robust_model}}{}
\addConstraint{\alpha + \frac{1}{(1-\text{$\beta$})}(\vecp^i)^\top\vecu^i \leq \theta,}{\quad}{i=1,\ldots,l}
\addConstraint{\vecu_{k}^i \geq -\vecw^\top\vecy_{k}^i - \alpha,}{\quad}{k=1,\ldots,S^i, i=1\ldots,l}
\addConstraint{\vecu_{k}^i \geq 0,}{\quad}{k=1,\ldots,S^i, i=1\ldots,l}
\addConstraint{\vecw^\top\vecytilde^i\geq\phi,}{\quad}{ i=1,\ldots,l}
\addConstraint{\vece^\top\vecw = w_0}{\quad}{}
\addConstraint{\vecw \in \mathbf{W}}{\quad}{}
\end{mini}

\chapter{Experiments and Results}
\label{sec:results} 

Having provided a solid explanation of the models we have selected to incorporate into the simulation analysis, the next step would be to evaluate their in-sample and out-of-sample performance, across one empirical dataset of daily returns, using certain performance criteria. To assess the magnitude of the potential gains that can actually be realized by an investor, it is necessary to analyze the \textit{out-of-sample} performance of the strategies from the optimizing models.~Afterwards, the performance of the non-robust portfolio models would be contradicted with their robust variants. The non-robust models considered along with their robust variants are shown in Table \ref{tab:models}.

\begin{table}
\fontsize{10pt}{9.0pt}\selectfont
\centering 
\caption{List of asset-allocation models} 
\begin{small}
\begin{tabular}{@{}lrrrr@{}}\toprule
    {Model Classification} & & & {Abbreviation}  \\ \midrule
   \textbf{Non-Robust Models}  &  \\ 
    Mean Variance   & & & \eqref{first_traditional_model}  \\
    Omega ratio   & & & \eqref{fourth_traditional_model}  \\ 
    CVaR   & & & \eqref{Rocka}  \\\midrule
    \textbf{Robust Models}   &   \\
    Robust Mean Variance   & & & \eqref{first_robust_model}  \\
    Robust Ellipsoidal   & & &  \eqref{second_robust_model}  \\
    Robust Multi-objective  & & & \eqref{third_robust_model}  \\
    Worst-case Omega   & & & \eqref{fourth_robust_model}  \\
    Worst-case CVaR   & & & \eqref{fifth_robust_model}  \\\midrule
   \label{tab:models}   
\end{tabular}
\end{small}
\end{table}

\section{Methodology}
\label{sec:methodology} 

We perform simulations based on historical data publicly available from Yahoo finance, acquired from time series for the index S \& P 500 spanning the period from January 1, 2005, to December 31, 2016. S \& P 500 is the most common equity index and is often used as a benchmark for the developed equities~\cite{kapsos2014worst}. This dataset comprises approximately 500 stocks from the New York market. For this analysis, we use 20 portfolios to optimize. This intriguing period would be a challenging test-bed for the framework considered, since it incorporates the year, where the global financial crisis took place in 2008~\cite{reavis2012global}, leading to a highly unpredictable factor in terms of the performance of the models considered.

\subsection{Rolling Windows Approach}
\label{sec:rolling} 

\begin{figure}[!h]		
  \centering
   \vspace{-0.5cm}
   \includegraphics[width=1.1\textwidth]{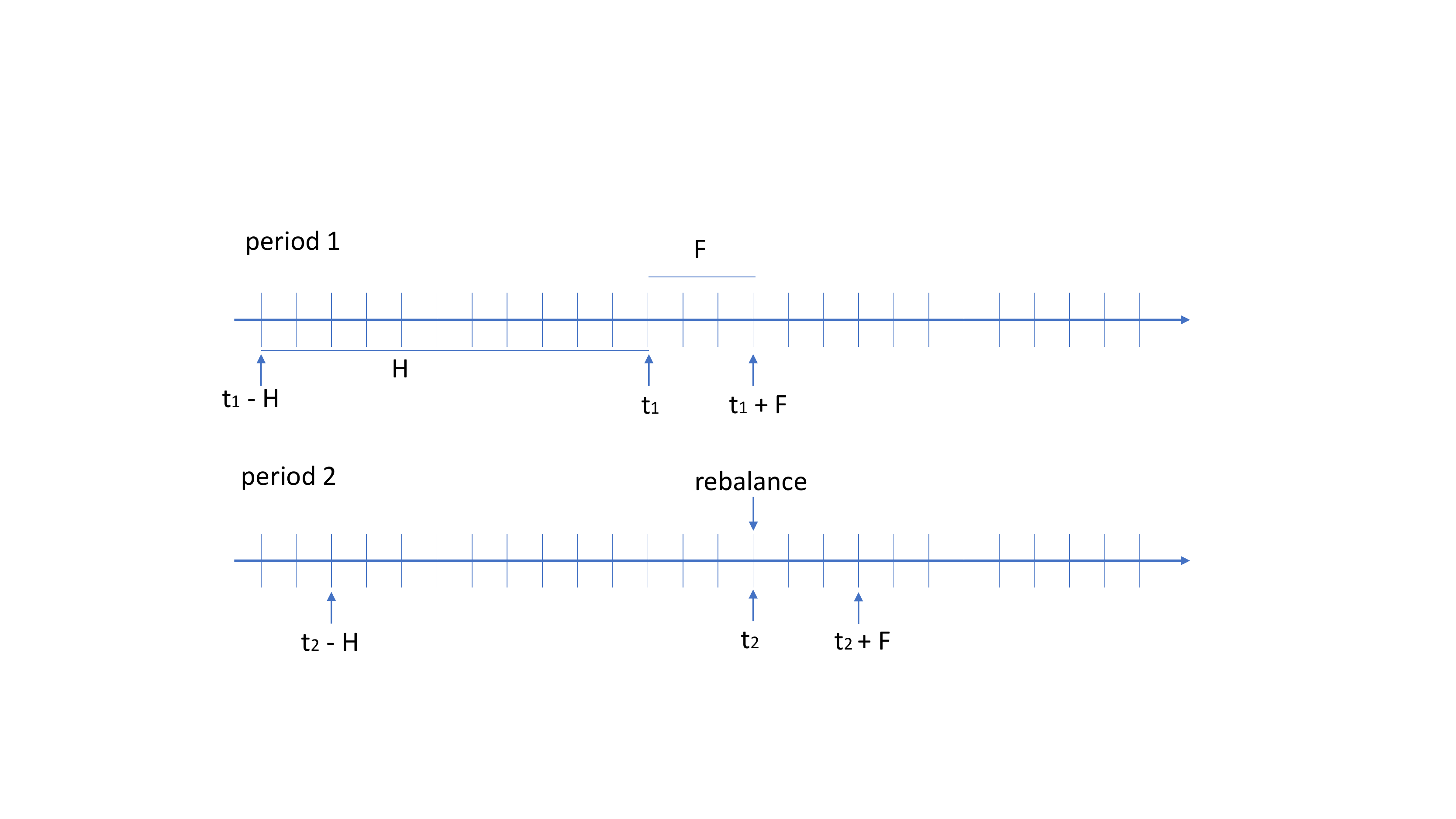}
  \vspace{-2cm}
   \caption{Illustration of rolling windows optimization}
  \label{rolling_windows}
\end{figure}

Inspired by the procedure proposed by Gilli and Schumann~\cite{gilli2011optimal},we conduct rolling-window backtests with a historical window of length H, and an out-of-sample holding period of length F. We set H to $250$ business days, F to $63$ business days. Thus we optimize at point in time $t_1$ on data from $t_1-H$ to $t_1-1$, the resulting portfolio is held until $t_2 = t_1+F$. At this point, a new optimal portfolio is computed, using data from $t_2-H$ until $t_2-1$, and the existing portfolio is rebalanced. This new portfolio is then held until $t_3 = t_2+F$, and so on. This is illustrated in Figure \ref{rolling_windows} for the first two periods. With our dataset, we have exactly 44 investment periods. From each run, we estimate the parameters needed to implement a particular strategy. These estimated parameters are then used to determine the portfolio weights for the assets. We optimize the first time in January 2005 $(t_1)$, the last date being 31 December 2016 $(t_{44})$. The outcome of this rolling-window approach is a series of $t_n-H$ monthly out-of-sample returns, where $t_n$ denotes the length of the rolling window approach.

\begin{itemize}
\item
Estimation period: One-year rolling window
\item
Test period: One quarter rolling window
\end{itemize}

Therefore, from the historical window in every period, the simulation incorporates four quarters of each year (in-sample) and one quarter (out-of-sample), where the quarters do not necessarily originate from the beginning of each year; each run could start from the middle of each year, but the spanning period will always be one year. Based on the outputs acquired from the in-sample period, the simulation uses these data to evaluate the efficiency of the portfolios created in the out-of-sample period.
Our goal is to study the performance of the aforementioned models based on the data acquired from the international portfolio market of New York.

At this point, we make a clarification about the procedure adopted exclusively for the Worst-case Omega ratio \eqref{fourth_robust_model} and Worst-case CVaR \eqref{fifth_robust_model}. For the nominal Omega ratio \eqref{fourth_traditional_model} and the nominal CVaR \eqref{Rocka}, the one year empirical distribution is used as input, whereas for the Worst-case Omega ratio \eqref{fourth_robust_model} and Worst-case CVaR \eqref{fifth_robust_model} four additional empirical distributions are provided by partitioning the one year period into four quarters of a year sub-periods. Note that for $l = 1$ the corresponding formulations in Sections \ref{sec:worst_omega_ratio} and \ref{sec:worst_cvar} are equivalent to their non-robust variants.~In terms of \eqref{third_robust_model}, the parameter $\varepsilon$ depicted in \eqref{robust_mult},  was computed as the 95 \% percentile of the bootstrap procedure by performing 1000 statistical computations, assuming that the respective sample required for a specific period (approximately 60 days), is a subset of the historical returns of the preceding year.~All in all, we consider the same amount of historical data, as pointed out earlier on in this Section. 

\section{Portfolio Performance Indicators}
\label{sec:metrics_for_performance}

Taking into account the time series of daily out-of-sample returns generated by each of the optimized models for the dataset S \& P 500, we compute the following quantities.

In order to assess the performance of the above simulation framework, for each of the optimization models used, we are going to use some metrics. First of all, we compute the mean return of model $k$, which is defined as the product of the average returns of the stocks in the portfolios $\bar{r}_k$ with the weights of the portfolios obtained from the simulation $w_k$.~Hence,

\begin{equation}
\hat{\mu}_k = \bar{r}_kw_k
\end{equation}

\noindent In addition, we calculate the standard deviation of portfolio returns of model $k$, which is defined as:

\begin{equation}
\hat{\sigma}_k = \sqrt{w_k^{\top}\Sigma_kw_k}
\end{equation}

\noindent where $\Sigma_k$ denotes the covariance matrix of stock returns.\\

Additionally, we compute the Sharpe ratio of model $k$, which stands for the sample mean of excess returns (over the risk-free rate) $\hat{\mu}_k$ divided by their sample standard deviation, $\hat{\sigma}_k$:

\begin{equation}
\text{SR}_k = \frac{\hat{\mu}_k}{\hat{\sigma}_k}
\end{equation} 

\noindent We also include the Sortino ratio of model $k$, which is defined as the ratio of mean returns of model $k$ with the standard deviation of negative returns for this model.~Thus,

\begin{equation}
\label{Sortino_computation}
\text{SoR}_k = \frac{\hat{\mu}_k}{S[\text{max}(0,-Rw_k)]}
\end{equation} 

\noindent where $R$ signifies the return data and $S$[] denotes the calculation of the standard deviation of model $k$.

\noindent Another instance we want to capture is the Omega ratio of model $k$. For this metric, we employ the quantity \eqref{type_for_Omega} in Section \ref{sec:omega}, fed with the weights which we acquire from the simulation horizon of model $k$.

\noindent  In a similar manner to Omega ratio, we compute the Conditional Value at Risk~ (CVaR) for each model $k$. For this metric we employ the quantity \eqref{Rocka} in Section \ref{sec:cvar}, adjusted with the respective weights we obtain from the simulation data for each model $k$.

Those computations are being performed for both the in-sample data and the out-of-sample data of the simulation procedure explained earlier on.

\section{Composition of Portfolios}
\label{sec:composition}

Except for the above risk-return performance measures, the composition of the portfolios is also analyzed. The characteristics of a portfolio's composition relate to management issues, such as the monitoring and rebalancing of the portfolio, as well as its management~(transaction) costs. To this end, first, we consider the number of assets in the portfolios constructed by each model.

\noindent To get a sense of the amount of trading required to implement each portfolio strategy, we compute the portfolio turnover, defined as the sum of the absolute value of the trades across the N variable assets~\cite{demiguel2007optimal}. Hence,

\begin{equation}
\text{Turnover} = \sum_{j=1}^{N}\left(|\vecwhat_{k,j,t+1}-\vecwhat_{k,j,t}|\right)
\end{equation}

\noindent in which $\vecwhat_{k,j,t}$ is the portfolio weight of asset $j$ at time $t$ under each respective optimized model $k$. The turnover quantity defined above can be interpreted as the average percentage of wealth traded in each period.~Turnover refers to the sum of absolute differences in portfolio weights compared to the previous quarter.

An additional feature we want to capture is the average diversification index from all the periods for every single run. As explained by Kim et al.~\cite{kim2013robust} one of the shortcomings of the mean-variance model is its tendency to put much weight on only a few assets. Having this in mind, if robust portfolios consist of more assets, the higher correlation with fundamental factors that we observe could be due to diversification. Blume and Freund~\cite{van1975asset} introduced a portfolio diversification measure: the deviation of a portfolio from the market portfolio. Since the weight of each security in the market portfolio would be very small, they proceeded to an approximation scheme with the sum of the squares of the proportions invested in each stock.

\begin{equation}
\text{Diversification} = \sum_{j=1}^{N}\left(\vecw_{k,j,t}-\vecw_{m}\right)^2 = \sum_{j=1}^{N}\left(\vecw_{k,j,t}-\frac{1}{N_m}\right)^2 \approx \sum_{j=1}^{N}\vecw_{k,j,t}^2
\end{equation}

\noindent where $N$ is the number of stocks in the portfolio $N_m$ is the number of stocks in the market portfolio,  and $\vecw_{m}$ is the weight given to a security in the market portfolio.

In terms of the diversification index, we can see in Table \ref{tab:metrics}  that among the three extensions of the mean-variance framework in the robust sense, comprising \eqref{first_robust_model}, \eqref{second_robust_model} and \eqref{third_robust_model}, only \eqref{first_robust_model} performs worse than the \eqref{first_traditional_model}. Moreover, \eqref{fourth_traditional_model} and \eqref{Rocka} attain a lower level of diversification compared to their robust variants \eqref{fourth_robust_model}, \eqref{fifth_robust_model} respectively. It is rather difficult to conclude that robust portfolios are more diversified than mean-variance portfolios. This makes the task of determining, whether the robust models are systematically superior to the non-robust models used in portfolio optimization more challenging.

\begin{table}[!htb]
\caption{Descriptive statistics for the composition of the portfolios}
\resizebox{\textwidth}{!}{%
\begin{tabular}{@{}lrrrrrrrrr@{}}\toprule
   & {$\eqref{first_traditional_model}$} & {$\eqref{fourth_traditional_model}$} & {$\eqref{Rocka}$} & {$\eqref{first_robust_model}$} & {$\eqref{second_robust_model}$} & {$\eqref{third_robust_model}$} & {$\eqref{fourth_robust_model}$} & {$\eqref{fifth_robust_model}$} \\ \midrule
   \textbf{Assets in portfolio} & $17.1591$ & $16.9773$ & $11.5375$ & $80.4636$ & $147.8466$ & $363.1795$ & $15.8182$ & $12.7602$ \\ 
   \textbf{Diversification index} & $0.2244$ & $0.1755$ & $0.2571$ & $0.3303$ & $0.0618$ & $0.0029$ & $0.1691$ & $0.2020$ \\
   \textbf{Turnover}  & $0.1540$ & $0.1489$ & $0.1719$ & $0.3809$ & $0.0709$ & $0.0139$ & $0.1466$ & $0.1667$ \\ \midrule
   \label{tab:metrics}   
\end{tabular}}
\end{table}

\noindent Based on Table \ref{tab:metrics} , we detect that for the three robust models \eqref{first_robust_model},\eqref{second_robust_model}\\and~\eqref{third_robust_model} there is a significant increase between the number of stocks in the portfolios~(which is not desirable), compared to the non-robust model \eqref{first_traditional_model}. This isn’t the case, however with the remaining two robust models \eqref{fourth_robust_model},\eqref{fifth_robust_model}. More specifically, the portfolios developed with the \eqref{fourth_robust_model} model have a slightly lower number of assets compared to the portfolios developed with~\eqref{fourth_traditional_model}.~On the other hand, for this metric explicitly, we note that \eqref{Rocka} performs slightly better compared to the robust counterpart~\eqref{fifth_robust_model}.

\begin{figure}[!htb]	
  \includegraphics[width=\textwidth]{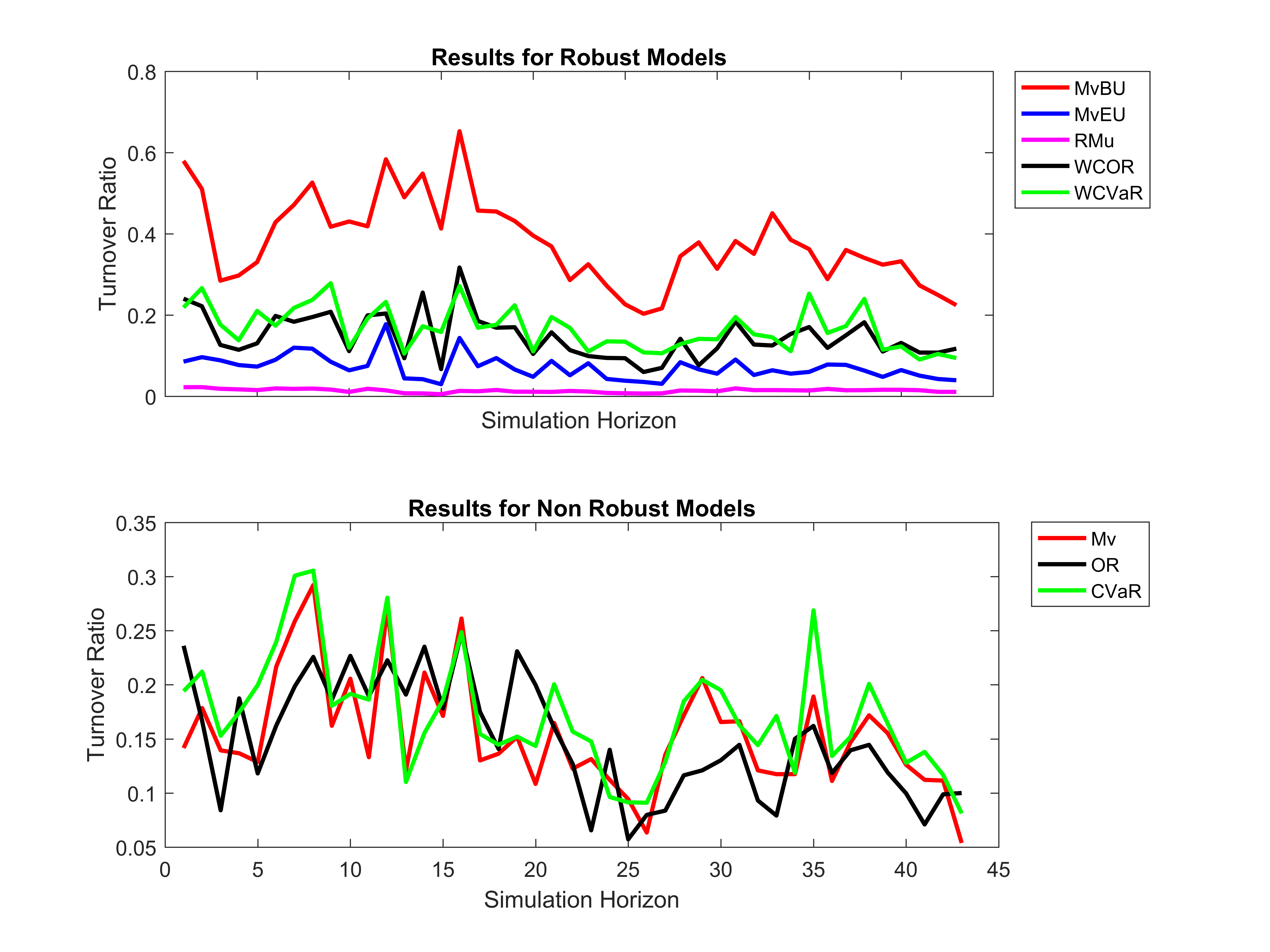}
   \caption{Turnover ratio for the considered strategies}
  \label{turnover}
\end{figure}

The turnover ratio is illustrated in Figure \ref{turnover} for the non-robust models along with their robust counterparts across the simulation horizon.~Comparing the portfolio turnover for the different models, we see that the turnover for the robust variant of the sample-based mean-variance portfolio, equipped with the box uncertainty \eqref{first_robust_model} is greater than the rest of the models. A general finding suggests that with the exception of \eqref{first_robust_model}, each other robust model achieves systematically a lower turnover than its non-robust counterpart. Moreover, it is interesting to note a peak value for the robust mean-variance model somewhere around the $16^{th}$ period of the simulation, which coincides with the year 2008, when the collapse of the U.S. housing market triggered the financial crisis, leading to dramatic plunge of major stock markets~\cite{reavis2012global}. This is an indicator, that the simulation procedure can accurately replicate the incident of the financial crisis carried out in 2008.

\newpage

\section{Portfolio Performance Results}
\label{sec:computation}

In this Section, we report the results of the computational experiments with the robust portfolio selection framework proposed in the Thesis. The objective of these computational experiments was to contrast the performance of the non-robust portfolio selection strategies with that of the robust portfolio selection strategies. The purpose of these experiments was to focus on the benefit accrued from robustness; All the computations were performed in MATLAB~(R2016b) using the Gurobi solver.

There are certain features we want to capture through this experimental procedure. Opting for a more concrete interpretation of the results obtained from all the corresponding simulation periods, the results shown hereafter were averaged over each run for all efficient portfolios derived through each model (20 portfolios for each model). Furthermore, we give the in-sample and out-of-sample performance of all the strategies considered in the simulation in an average form spanning all the horizon. To assess the magnitude of the potential gains that can actually be realized by an investor, it is necessary to analyze the out-of-sample performance of the strategies from the optimization models~\cite{demiguel2007optimal}.

\subsection{Comparison of non-robust models}
\label{sec:comparison_non_robust}

Taking a look at the mean return accumulated from each model, we can realize that for the non-robust models, there is a difference between the values obtained for the in-sample data and for the out-of-sample data.\\
Overall, the~\eqref{fourth_traditional_model} model does perform in a superior manner compared to \eqref{first_traditional_model} and \eqref{Rocka} for each performance indicator we have incorporated in Section \ref{sec:metrics_for_performance}, in terms of the out-of-sample data~(except the mean return, where these models perform equally). Regarding the Conditional Value at Risk metric, three different confidence levels are employed. We start with a 90\% confidence level in the first case and then we augment it to 95\% in the second case reaching 99\% in the third case. We notice, that even though the \eqref{Rocka} model should possess a higher value for each one of the different confidence levels imposed, for the Conditional Value at Risk metric, this model does attain nevertheless the worst value among \eqref{first_traditional_model} and \eqref{fourth_traditional_model} for a 99\% confidence level, in terms of the out-of-sample data, that being 0.0302 compared to 0.247 and 0.0285, for  \eqref{fourth_traditional_model} and \eqref{first_traditional_model}, respectively. Furthermore, we can detect that the results we get for the standard deviation and for the Conditional Value at Risk (along with 3 confidence levels) are quite consistent with respect to the in-sample data compared to the out-of-sample data, for these non-robust models, with almost indistinguishable discrepancies. The detailed results for the non-robust models are shown in Table \ref{tab:conclusive_non_robust}.

\begin{table}[!htb]
\fontsize{10pt}{9.0pt}\selectfont
\centering 
\caption{Performance metrics for the non-robust models}
\begin{tabular}{@{}lrrrr@{}}\toprule
   & {$\eqref{first_traditional_model}$} & {$\eqref{fourth_traditional_model}$} & {$\eqref{Rocka}$} \\ \midrule
  \textbf{Mean return~(in-sample)} & $0.0017$ & $0.0017$ & $0.0018$ \\
     \textbf{Mean return~(out-of-sample)} & $0.0003$ & $0.0003$ & $0.0003$ \\ \midrule
   \textbf{Standard deviation~(in-sample)} & $0.0093$ & $0.0080$ & $0.0101$ \\  
       \textbf{Standard deviation~(out-of-sample)} & $0.0105$ & $0.0093$ & $0.0112$	 \\  \midrule
    \textbf{Sharpe ratio~(in-sample)} & $0.2119$ & $0.2593$ & $0.2032$ \\ 
       \textbf{Sharpe ratio~(out-of-sample)} & $0.0649$ & $0.0694$ & $0.0584$ \\ \midrule
   \textbf{Sortino ratio~(in-sample)} & $0.4064$ & $0.5148$ & $0.4224$ \\ 
         \textbf{Sortino ratio~(out-of-sample)} & $0.1331$ & $0.1391$ & $0.1218$ \\ \midrule
    \textbf{Omega ratio~(in-sample)} & $1.7797$ & $2.0802$ & $1.7288$  \\ 
   \textbf{Omega ratio~(out-of-sample)} & $1.2488$ & $1.2624$ & $1.2262$  \\ \midrule
          \textbf{90\% CVaR~(in-sample) (\%)} & $0.0150$ & $0.0127$ & $0.0153$  \\ 
   \textbf{90\% CVaR~(out-of-sample) (\%)} & $0.0185$ & $0.0161$ & $0.0197$  \\ \midrule
   \textbf{95\% CVaR~(in-sample) (\%)} & $0.0188$ & $0.0161$ & $0.0182$  \\ 
   \textbf{95\% CVaR~(out-of-sample) (\%)} & $0.0226$ & $0.0196$ & $0.0239$  \\ \midrule
   \textbf{99\% CVaR~(in-sample) (\%)} & $0.0271$ & $0.0229$ & $0.0268$  \\ 
   \textbf{99\% CVaR~(out-of-sample) (\%)} & $0.0285$ & $0.0247$ & $0.0302$  \\ \midrule
   \label{tab:conclusive_non_robust}
\end{tabular}
\end{table}

\subsection{Comparison of robust models}
\label{sec:comparison_robust}

In terms of the robust variants for the computation of mean-return, all models perform better in-sample rather than out-of-sample. Among the in-sample Sharpe ratios for the robust models at hand, \eqref{first_robust_model} attains the highest value (0.4027) for the in-sample data, implying that this model can perform in a satisfying manner within the bounds of the uncertainty~(''box'') imposed. We see, that the Conditional Value at Risk metric we get explicitly from the optimization of~\eqref{fifth_robust_model} doesn't attain the best value among the robust models for none of the confidence levels considered, as far as the in-sample data are concerned.~An instance, which comes as a surprise is, that for the out-of-sample data,~\eqref{fifth_robust_model} model does acquire the second worst value in terms of Conditional Value at Risk metric among the robust models for each of the confidence levels mentioned. This behaviour could be attributed to the fact, that the value of~\eqref{fifth_robust_model} is actually the average of the 20 portfolios considered, where each one corresponds to a certain risk and return, depending on its location in the efficient frontier curve. Indeed, one portfolio does minimize the risk, but since we are interested in the average form of the risk, it is possible to get an inferior value for the Conditional Value at Risk metric for model~\eqref{fifth_robust_model} in comparison with the other robust models. Additionally, another factor which could cause this behaviour lies in the fact of the step imposed for each robust model, to construct the efficient frontier. With the exception of \eqref{fourth_robust_model}, which represents a single portfolio, every other robust model used an alternative step to designate the efficient frontier. In a respective manner,~\eqref{fourth_robust_model} does acquire the worst value as far as the Omega ratio metric is concerned for the out-of-sample data, among the rest of the robust models. The fact, in this case, is that as was mentioned before,~\eqref{fourth_robust_model} poses a sole portfolio, so no average form was taken and nevertheless, it attained a lesser value (1.2351) in comparison with the other robust models, which used an average form to account for the Omega ratio metric.

Another finding which is suggested by the results shown in Table \ref{tab:conclusive_robust}\footnote[3]{The in-sample values for ~\eqref{first_robust_model} with respect to the Sortino ratio and Omega ratio, were not left intentionally blank. This model had an extremely low standard deviation of negative returns (nearly 0) and so the computation in \eqref{Sortino_computation} was infeasible. For the Omega ratio, the daily returns for the in-sample data were always positive and the computation of \eqref{type_for_Omega} measures the ratio of gains to losses. Considering that there are no losses for \eqref{first_robust_model} we couldn't get a value for this metric either.}, is that \eqref{third_robust_model} performs in quite close proximity, not only for the in-sample data but also for the out-of-sample data, as well.

\begin{table}[!htb]
\fontsize{10pt}{9.0pt}\selectfont
\centering 
\caption{Performance metrics for the robust models}
\begin{tabular}{@{}lrrrrrr@{}}\toprule
   & {$\eqref{first_robust_model}$} & {$\eqref{second_robust_model}$} & {$\eqref{third_robust_model}$} & {$\eqref{fourth_robust_model}$} & {$\eqref{fifth_robust_model}$} \\ \midrule
  \textbf{Mean return~(in-sample)}   & $0.0013$ & $0.0007$ & $0.0007$ & $0.0016$ & $0.0015$ \\
     \textbf{Mean return~(out-of-sample)}  & $0.0003$ & $0.0004$ & $0.0005$ & $0.0002$ & $0.0004$ \\ \midrule
   \textbf{Standard deviation~(in-sample)}  & $0.0054$ & $0.0067$ & $0.0114$ & $0.0086$ & $0.0091$ \\  
       \textbf{Standard deviation~(out-of-sample)}  & $0.0086$ & $0.0073$ & $0.0111$ & $0.0093$ & $0.0099$ \\  \midrule
    \textbf{Sharpe ratio~(in-sample)}  & $0.4027$ & $0.1268$ & $0.0772$ & $0.2396$ & $0.1952$ \\ 
       \textbf{Sharpe ratio~(out-of-sample)}  & $0.0661$ & $0.0880$ & $0.0704$ & $0.0601$ & $0.0702$ \\ \midrule
   \textbf{Sortino ratio~(in-sample)}  & $-$ & $0.2229$ & $0.1312$ & $0.4733$ & $0.4002$ \\ 
         \textbf{Sortino ratio~(out-of-sample)}  & $0.1378$ & $0.1723$ & $0.1303$ & $0.1243$ & $0.1418$ \\ \midrule
    \textbf{Omega ratio~(in-sample)}  & $-$ & $1.4210$ & $1.2413$ & $1.9592$ & $1.6885$ \\ 
   \textbf{Omega ratio~(out-of-sample)}  & $1.2633$ & $1.3205$ & $1.2507$ & $1.2351$ & $1.2667$ \\ \midrule
          \textbf{90\% CVaR~(in-sample) (\%)}  & $0.083$ & $0.0119$ & $0.0210$ & $0.0137$ & $0.0143$ \\ 
   \textbf{90\% CVaR~(out-of-sample) (\%)}  & $0.0149$ & $0.0128$ & $0.0197$ & $0.0165$ & $0.0173$ \\ \midrule
   \textbf{95\% CVaR~(in-sample) (\%)}  & $0.0104$ & $0.0149$ & $0.0260$ & $0.0172$ & $0.0170$ \\ 
   \textbf{95\% CVaR~(out-of-sample) (\%)}  & $0.0181$ & $0.0155$ & $0.0236$ & $0.0199$ & $0.0211$ \\ \midrule
   \textbf{99\% CVaR~(in-sample) (\%)}  & $0.0151$ & $0.0220$ & $0.0367$ & $0.0243$ & $0.0224$ \\ 
   \textbf{99\% CVaR~(out-of-sample) (\%)} & $0.0230$ & $0.0194$ & $0.0289$ & $0.0253$ & $0.0264$ \\ \midrule
   \label{tab:conclusive_robust}
\end{tabular}
\end{table}

\newpage

\subsection{Robust versus non-robust models}
\label{sec:robust_versun_non}

We notice that for all the models~(robust and non-robust) with the exception of~\eqref{third_robust_model}, there is a smaller standard deviation for the in-sample data, rather than for the out-of-sample data. 

From the results presented in Tables \ref{tab:conclusive_non_robust} and \ref{tab:conclusive_robust}, it is evident that all models perform considerably better for the in-sample tests compared to the out-of-sample ones, in terms of the Sharpe ratio. Although this behaviour verifies the well-known weaknesses of using classical sample-based estimates of the moments of asset returns to implement Markowitz's mean-variance portfolios, the difference between the in-sample Sharpe ratio (0.0772) and out-of-sample Sharpe ratio (0.0704) for \eqref{third_robust_model} is only marginal. This could be attributed to the joint uncertainty set established to construct the robust multi-objective model \eqref{third_robust_model}, rendering it more resilient to uncertainty shortcomings.

Additionally, non-robust models perform much better in terms of the Sortino ratio for the in-sample data in contrast with the out-of-sample data. This applies, as well to the robust models.

We shift our attention to the Omega ratio. As was expected for the in-sample data \eqref{fourth_robust_model} attains the highest value among the robust models (1.9592), as well as \eqref{fourth_traditional_model} does  among the non-robust models (2.082). Moreover, it is evident that the non-robust model \eqref{fourth_traditional_model} performs slightly better than the robust model \eqref{fourth_robust_model} in terms of in-sample data.~On the contrary, taking a closer look in the out-of-sample data, it is not a trivial task to pinpoint why \eqref{fourth_robust_model} performs in the worst style, attaining a value (1.2351) among  the rest of the robust models, considering that the Omega ratio we should get for \eqref{fourth_robust_model} stems from the optimization process of \eqref{fourth_robust_model} and should theoretically give us the highest value.

As reported in Table \ref{tab:conclusive_non_robust} the values for the out-of-sample Conditional Value at Risk metric show a steady rise as the confidence increases, not only for the non-robust models but also for the robust models as well. 

Aiming to present a more concrete realization of the behaviour of the robust models in comparison with their non-robust variants, we consider the following Table.~Here, we state in which metrics, do the non-robust models acquire better values than their robust equivalents.~We realize, that robust optimization techniques would not always yield better performance.~(RO) tries to designate the best strategy, using historical data, however, the prediction cannot always be accurate and errors might arise through this procedure.

\begin{table}[!ht]
\caption{Efficiency of Robust Extensions of the models}
   \centering
   \begin{tabular}{llllll}
   \toprule
   & {\bfseries \quad Mv} & {\bfseries OR} & {\bfseries CVaR}  \\
   \midrule
   {\bfseries MvBU} & \quad \quad - & \cellcolor[gray]{0.9}  & \cellcolor[gray]{0.9} 
      &  &  \\ \midrule
   {\bfseries MvEU} & \quad \quad -  & \cellcolor[gray]{0.9} & \cellcolor[gray]{0.9} & 
      & \\ \midrule
      {\bfseries RMu} & Standard deviation & \cellcolor[gray]{0.9} & \cellcolor[gray]{0.9} & 
      & \\
      & Sortino ratio & \cellcolor[gray]{0.9} & \cellcolor[gray]{0.9} &  & 
      \\
   &90\% CVaR & \cellcolor[gray]{0.9} & \cellcolor[gray]{0.9} &  & 
      \\
     &95\% CVaR & \cellcolor[gray]{0.9} & \cellcolor[gray]{0.9} &  & 
      \\
     &99\% CVaR & \cellcolor[gray]{0.9} & \cellcolor[gray]{0.9} &  & 
      \\ \midrule
      {\bfseries WCOR} &\cellcolor[gray]{0.9}  &Mean return & \cellcolor[gray]{0.9}  & 
      & \\
      &\cellcolor[gray]{0.9}  &Sharpe ratio & \cellcolor[gray]{0.9}  & 
      & \\
      &\cellcolor[gray]{0.9}  &Sortino & \cellcolor[gray]{0.9}  & 
      & \\
      &\cellcolor[gray]{0.9}  &90,95,99\% CVaR & \cellcolor[gray]{0.9}  & 
      & \\ \midrule
      {\bfseries WCVaR} & \cellcolor[gray]{0.9} & \cellcolor[gray]{0.9} & \quad - & 
      & \\
   \bottomrule
   \end{tabular} 
\end{table}

Keeping in mind these explanations, we can rationally assume, that the majority of the results we obtain from the robust optimization framework are overall superior from those we get from classical optimization techniques, without this meaning that robust optimization is the only efficient way of handling problems of this architecture. In some instances, as the ones examined in the Thesis, there are metrics for which non-robust models perform better than robust models for the out-of-sample data.

\section{Validation of the Uncertainty Sets}
\label{sec:robustness_checks}

During the process of the evaluation procedure, certain assumptions were made, regarding the range and the architecture of the uncertainty sets employed to reach a tractable solution.~In this Section, we present the results regarding the uncertainty sets employed for each model and tests and whether the out-of-sample data are in accordance with the uncertainty sets formulated based on historical data~(in-sample).

Model \eqref{first_robust_model} assumes that the unknown future (\textit{out-of-sample}) mean return $\mu_i^f$ of stock $i$ will be such that $|\mu_i^f - \hat{\mu_i}| \leq \delta_i$, where $\delta_i$ was introduced previously in Section \ref{sec:uncer_estim} and $\hat{\mu_i}$ is the mean return of the stock according to the \textit{in-sample} data.~Our aim hereafter is to investigate whether the true mean return of the stocks calculated from the out-of-sample data does indeed satisfy this assumption. Our aim is to measure the frequency, where this condition is verified, within the considered bounds.

Model \eqref{second_robust_model} assumes that $(\mu - \hat{\mu})^T\varSigma_{\mu}^{-1}(\mu - \hat{\mu}) \leq \delta^2$, where $\delta$ was introduced as well in Section \ref{sec:counterpart} with $\hat{\mu}$ being the vector of the mean stock returns from in-sample data and $\mu$ the vector of out-of-sample mean returns of the stocks and $\varSigma_{\mu}$ defined using the in-sample data. As in the previous model, we test whether this assumption holds or not for each of the tests performed during the examined time period.

Model \eqref{third_robust_model} assumes that $\|\mu - \hat{\mu}\|+c\|\varSigma - \hat{\varSigma}\| \leq \varepsilon$ where parameter $\varepsilon$ signifies the boundary under which lie the 950 sorted random values of the bootstrap procedure with respect to the distribution followed by $\|\mu - \hat{\mu}\|+c\|\varSigma - \hat{\varSigma}\|$ for the in-sample data, mentioned in Section \ref{sec:rolling} and setting $c = 1$ as explained in Section \ref{sec:multiobjective}.

Model \eqref{fourth_robust_model} assumes that with multiple estimates for the omega ratio $\{\Omega_1, \Omega_2, \Omega_3, \Omega_4\}$, the best portfolio is the one that maximizes the worst of the omegas. As was formulated in the Thesis, Worst-case Omega ratio was solved with 4 mixtures , each corresponding to 4 quarters prior to the current quarter T, to obtain an optimal robust portfolio. We calculate the corresponding omegas of the optimal portfolio by $\Omega_1,\ldots,\Omega_4$ for each of the past 4 quarters and also compute the omega ratio $\Omega_T^R$ of the optimal robust portfolio for the out-of-sample quarter T, according to \eqref{type_for_Omega}. Moreover, we solve non-robust \eqref{fourth_traditional_model} and calculate its Omega ratio $\Omega_T$ for the out-of-sample quarter T. Then,

\begin{itemize}
\item
The robustification can be considered as \enquote{fully successful} if~ $\Omega_T^R \geq\\ \min\{\Omega_1, \Omega_2, \Omega_3, \Omega_4\}$ and $\Omega_T < \min\{\Omega_1, \Omega_2, \Omega_3, \Omega_4\}$
\item
The robustification can be considered as \enquote{partially successful} if~ $\Omega_T^R > \Omega_T$ 
\item
The robustification can be considered as \enquote{totally unsuccessful} if~ $\Omega_T^R < \min\{\Omega_1, \Omega_2, \Omega_3, \Omega_4\}$ and $\Omega_T \geq \min\{\Omega_1, \Omega_2, \Omega_3, \Omega_4\}$
\item
The robustification can be considered as \enquote{partially unsuccessful} if~ $\Omega_T^R < \Omega_T$ 
\end{itemize}

\noindent A similar approach is employed to compare the \eqref{fifth_robust_model} model to its nominal (non-robust) counterpart \eqref{Rocka}. In particular, let $\text{CVaR}_1, \ldots, \text{CVaR}_4$ denote the last years quarterly CVaRs used to derive the worst-case CVaR portfolio with the the \eqref{fifth_robust_model} model. The out-of-sample CVaR for the corresponding portfolio is $\text{CVaR}_T^R$, whereas the out-of-sample CVaR of the portfolio derived from the nominal \eqref{Rocka} model is denoted by $\text{CVaR}_T$. Then:

\begin{itemize}
\item
The robustification can be considered as \enquote{fully successful} if~ $\text{CVaR}_T^R \leq \max\{\text{CVaR}_1, \text{CVaR}_2, \text{CVaR}_3, \text{CVaR}_4\}$ and\\ $\text{CVaR}_T > \max\{\text{CVaR}_1, \text{CVaR}_2, \text{CVaR}_3, \text{CVaR}_4\}$
\item
The robustification can be considered as \enquote{partially successful} if~ $\text{CVaR}_T^R < \text{CVaR}_T$ 
\item
The robustification can be considered as \enquote{totally unsuccessful} if~ $\text{CVaR}_T^R > \max\{\text{CVaR}_1, \text{CVaR}_2, \text{CVaR}_3, \text{CVaR}_4\}$ and\\ $\text{CVaR}_T \leq \max\{\text{CVaR}_1, \text{CVaR}_2, \text{CVaR}_3, \text{CVaR}_4\}$
\item
The robustification can be considered as \enquote{partially unsuccessful} if~ $\text{CVaR}_T^R > \text{CVaR}_T$ 
\end{itemize}

\noindent For the two latter evaluation checks;~\eqref{fourth_robust_model} and \eqref{fifth_robust_model}, we make an adjustment, which will account for the level of impact we assign to each of the statements. So, the adjustment lies within the concept of the weighted average, where we give a weight equal to 1 if the statement for each single run of the simulation is \enquote{totally successful}, 0.5 if the statement is \enquote{partially successful},  -0.5 if the statement is \enquote{partially unsuccessful} and  -1 if the statement is \enquote{totally unsuccessful}.~Considering that each of the statements is denoted as $C(i)$, with $i=1$ representing the first statement and $i=4$, the last statement, we can express mathematically the reward function for each one of the 44 periods of the simulation, each period representing $j$. Hence,

\begin{equation}
\label{kill_bill}
 \text{Gain}_j = 1 \times C(1)+0.5 \times C(2) -1 \times C(3) - 0.5 \times C(4)
\end{equation}

\noindent All in all, having estimated how frequent a specific condition holds, namely for \eqref{first_robust_model},\eqref{second_robust_model},\eqref{third_robust_model} and the level of impact that \eqref{fourth_robust_model} and \eqref{fifth_robust_model} pose according to \eqref{kill_bill}, we subsequently  take the average of these quantities from all the assets participating in the portfolios at each run, for each of the 44 periods considered. We present the outcomes in Table~\ref{tab:average_checks}.

\begin{table}
\caption{Performance Checks}
\resizebox{\textwidth}{!}{%
\begin{tabular}{@{}lrrrrrr@{}}\toprule
     & {$\eqref{first_robust_model}$} & {$\eqref{second_robust_model}$} & {$\eqref{third_robust_model}$} & {$\eqref{fourth_robust_model}$} & {$\eqref{fifth_robust_model}$} \\ \midrule
   \textbf{Simulation Outputs}  & $0.9527$ & $0.6591$ & $0.7045$ & $-0.1477$ & $-0.1250$  
   \label{tab:average_checks}
\end{tabular}}
\end{table}

The next step of the process is to interpret the effectiveness of these validation checks, by counting the number of time periods for which each condition holds, either in terms of  the bounded uncertainty for models \eqref{first_robust_model},\eqref{second_robust_model},\eqref{third_robust_model} or either in terms of the gain function employed for models \eqref{fourth_robust_model} and \eqref{fifth_robust_model}. Viewing Table~\ref{tab:average_checks}, we realize that \eqref{first_robust_model} verifies the uncertainty condition imposed in Section \ref{sec:uncer_estim} for nearly the 96 $\%$ of the simulation periods, an element which implies that this model does have an exceptional behaviour out-of-sample.

As far as \eqref{second_robust_model} model is concerned, we realize that it performs within the given bounds for almost the 66 $\%$ of the simulation runs.

\eqref{third_robust_model} performs within the given threshold $\varepsilon$ for nearly the 71 $\%$ of the simulation periods.

Models \eqref{fourth_robust_model} and \eqref{fifth_robust_model} attain nearly the same score, which is a negative one. This implies, that there is an inclination towards the satisfaction of the term \enquote{partial unsuccessfulness}, meaning that these models didn't perform in the desired manner for the out-of-sample data.

\chapter{Conclusions}
\label{sec:conclusions} 

\markboth{}{}

In summary, although robust models decrease the sensitivity in parameter estimation errors, it is not a trivial task to measure how successfully the proposed models achieve their goals under practical settings. The verdict from this comparison between robust and non-robust models is, that there seems to be an amelioration in the results we get, without that being universal.~Robust optimization models cannot always cope with the uncertainty in a convincing manner, that being their major limitation. However, as depicted from the results shown above, they do indeed present a satisfying performance in terms of the different uncertainty architectures imposed on the robust models. In general, the out-of-sample results with respect to the robust models are superior to those we get for the non-robust models.~\eqref{second_robust_model} seems to perform in a superior manner \textit{out-of-sample} for nearly all the metrics considered, in terms of the robust models.~\eqref{third_robust_model} presents the most consistent behaviour among the robust models, since the values attained from the metrics considered, for the in-sample and the out-of-sample data, present infinitesimal deviations.~Some models perform better than others judging by different metrics, but the whole picture is that the results are promising and pose the need for the investigation of more advanced techniques in the field of robust optimization. At a subsequent stage of the evaluation, the validation of the uncertainty sets was examined in Section \ref{sec:robustness_checks}, to check whether the robust models do indeed perform during the simulation runs, within their respective bounds.~Based on the results acquired, we can deduce that~\eqref{first_robust_model} presents an excellent performance, considering that it verifies the box uncertainty mentioned in Section \ref{sec:uncer_estim} for nearly the 96 $\%$ of the simulation periods.
Models~\eqref{second_robust_model} and~\eqref{third_robust_model} perform in a satisfying manner, verifying their specific uncertainty assumptions for the 66  $\%$ and 69  $\%$ of the simulation periods accordingly.
We considered these specific models, so as to capture the effects of \textit{robust optimization framework} under different mathematical representations of the models, each one tackling a different objective.~A next step of this procedure, could be to incorporate even more flexible models of data-driven uncertainty and not just of predetermined architecture, as in the proposed methodology perceived in the Thesis.~Another instance that could be examined, is the impact of \textit{robust optimization} for portfolio selection on industries with different investing policies.

\newpage{\pagestyle{empty}\cleardoublepage}

\bibliographystyle{abbrv}
\bibliography{thesis}

\markboth{}{}


\end{document}